\documentclass[fleqn,usenatbib]{mnras}

\usepackage{newtxtext,newtxmath}
\usepackage[T1]{fontenc}

\DeclareRobustCommand{\VAN}[3]{#2}
\let\VANthebibliography\thebibliography
\def\thebibliography{\DeclareRobustCommand{\VAN}[3]{##3}\VANthebibliography}

\usepackage{graphicx}	% Including figure files
\usepackage{amsmath}	% Advanced maths commands
\usepackage{amsfonts}
\usepackage{relsize}
\usepackage{mathtools} %big curly bracket
\usepackage{color,soul} %highlighting
\usepackage{bm}
\usepackage[dvipsnames]{xcolor}
\usepackage{rotating}
\usepackage{array, makecell}
\usepackage{tabularx}
\usepackage{colortbl}
\usepackage{hyperref}

\hypersetup{
    colorlinks=true,
    linkcolor=blue,
    filecolor=magenta,      
    urlcolor=blue,
    pdftitle={Overleaf Example},
    pdfpagemode=FullScreen,
    }

\usepackage{physics}

\newcommand{\TF}{Tully-Fisher }
\newcommand{\hi}{H\thinspace{\protect\scriptsize I }}

\title[Calibration of the WISE \TF relation]{Calibration of the \TF relation in the WISE W1 ($3.4\mu \rm m$) and W2 ($4.6\mu \rm m$) Bands}

\author[R. Bell et al.]
{\parbox{\textwidth}{Rianna Bell$^{1}$\thanks{E-mail: rianna.bell@uq.edu.au}, Khaled Said$^{1}$, Tamara Davis$^1$, T.H. Jarrett$^2$}
\vspace{0.4cm}\\
\parbox{\textwidth}{
$^{1}$School of Mathematics and Physics, The University of Queensland, Brisbane, QLD 4072, Australia\\
$^2$Astronomy Department,  University of Cape Town, Private Bag X3, Rondebosch, 7701, South Africa\\
}}
\date{Accepted XXX. Received YYY; in original form ZZZ}

\pubyear{\today}

\begin{document}
\label{firstpage}
\pagerange{\pageref{firstpage}--\pageref{lastpage}}
\maketitle

% Abstract of the paper
\begin{abstract}
In this paper, we present our calibrations of the \TF relation in the mid-infrared W1 ($3.4\mu$m) and W2 ($4.6\mu$m) bands, using large samples 848 galaxies and 857 galaxies from 31 clusters in the W1 and W2 bands respectively. In this calibration we performed two main corrections: a correction for the cluster population incompleteness bias, and a morphological type correction. The calibration was performed using a new, iterative bivariate fitting procedure. For these calibrations we used the total absolute magnitudes, and \hi linewidths $W_{F50}$ derived from the HI global profiles as a measure of the rotational velocities. We then performed two additional calibrations on the same sample using (i) the isophotal magnitudes and (ii) the average rotational velocities measured along the flat sections of the spatially resolved rotation curves of the galaxies, which were obtained from the empirical conversion between rotational velocity definitions. We compared these three calibrations to determine whether the use of isophotal magnitudes, or spatially resolved rotational velocities have a significant impact on the scatter around the TF relations in the W1 and W2 bands. We found that the original calibrations using total magnitudes and \hi linewidths had the smallest total scatters. These calibrations are given by $M_{\rm Tot, W1} = (2.02 \pm 0.44) - (10.08 \pm 0.17)\log_{10}(W_{F50})$ and $M_{\rm Tot, W2} = (2.00 \pm 0.44) - (10.11 \pm 0.17)\log_{10}(W_{F50})$, with associated total scatters of $\sigma_{W1} = 0.68$ and $\sigma_{W2} = 0.69$. Finally, we compared our calibrations in the mid-infrared bands with previous calibrations in the near-infrared J, H and K bands and the long-wavelength optical I band, which used the same two corrections. The differences between these relations can be explained by considering the different regions and components of spiral galaxies that are traced by the different wavelengths. The codes used for the calibration of the \TF relation are available from \url{ https://github.com/RiannaBell/WISE-TF-Calibration}. 
\end{abstract}

% Select between one and six entries from the list of approved keywords.
% Don't make up new ones.

%%%%%%%%%%%%%%%%% BODY OF PAPER %%%%%%%%%%%%%%%%%%
\section{Introduction}
The \TF (TF) relation is an empirical correlation between the rotational velocities and luminosities of spiral galaxies \citep{Tully_Fisher1977}. Since it was first observed, the TF relation has remained one of the most important \textit{distance-indicator relations} in modern cosmology, having been used to derive the distances to, and peculiar velocities of, thousands of spiral galaxies throughout the universe \citep{Strauss-Willick1995}.

Comparisons of the slope, intercept and scatter of the TF relation at different wavelengths can provide insight into aspects of galaxy evolution as different components of galaxies are formed at different stages of their evolution and produce emissions at different wavelengths.

However, for the TF relation to be used effectively, it must first be calibrated using spiral galaxies that have known distances. Calibration of the TF relation is a complex process which must overcome challenges posed when working with observational samples, including measurement errors and statistical biases. To date, the TF relation has been calibrated in almost all of the optical and near-infrared bands using a variety of techniques. 

Early calibrations of the TF relation primarily used small samples with apparent magnitudes measured in the optical bands. However, it was realised that the scatter around the TF relation could be reduced by using larger samples sizes with apparent magnitudes measured at longer wavelengths, which are less susceptible to dust extinction both in the Milky Way and in the galaxies being observed \citep{M_Aaronson1979}.

In particular, with the release of the The Two Micron All Sky Survey
\citep[2MASS;][]{M_Skrutskie2006}, which provided near-infrared apparent magnitude measurements for millions of galaxies, it was shown that the calibrated TF relation has a significantly reduced scatter in the near-infrared K-band, which has almost negligible extinction \citep{K_Masters2008}. This led to significant interest in extending the calibrations into the mid-infrared bands, with the hope of further improving the accuracy of the relation \citep{Freedman-Madore2010}.

Historically, there have been two main approaches to calibrating the TF relation at any wavelength. The first is to calibrate the TF relation using a bivariate fit, which treats neither the rotational velocity nor the the luminosity as the dependent variable. This type of approach will generally include a correction for the statistical biases present in the data (generally a cluster population incompleteness bias correction), as well as additional corrections to account for photometric effects, and sample variation. Of the previous calibrations in the mid-infrared bands, \cite{D_Lagattuta2013} used a bivariate fitting approach to calibrate the TF relation in the W1 band using a sample of 568 galaxies. However, this calibration did not correct for the statistical biases within the sample, only correcting for photometric effects and morphological variation within the sample. 

The second approach to calibrating the TF relation is to use an inverse fitting method in which the rotational velocity is treated as the dependent variable. This method is primarily used in order to mitigate the need for a statistical bias correction, as the inverse TF relation is less significantly impacted by statistical biases than either of the forward or bivariate fitting techniques. The inverse TF relation has been used by \cite{J_Sorce2013}, \cite{J_Neill2014} and \cite{E_Kourkchi2019} in the mid-infrared bands on samples consisting of 213, 310, and 584 galaxies respectively. Each of these calibrations also included a colour correction. However, although the use of the inverse TF relation has been shown to significantly reduce the impact of statistical biases on the slope and intercept of the TF relation, it does not produce a completely bias free calibration \citep{J_Willick1994}. This is particularly significant for the cosmological applications of the TF relation, as a biased slope and intercept may result in inaccurate distance measurements for galaxies. 

In addition to these different calibration approaches, historical calibrations of the TF relation across all wavelengths have differed in both the apparent magnitude definitions and the rotational velocity definitions used. Of the calibrations in the mid-infrared bands, both isophotal and total (extrapolated) apparent magnitudes have been previously used \citep{K_Masters2008, K_Said2015}. However, the impact of these different apparent magnitudes measurements on the calibrated TF relation and its associated scatter is unknown. Recently, \cite{F_Lelli2019} has discussed the impact of different rotational velocity measurements on the total scatter around the TF relation, showing that the use of the average circular velocity along the flat region of spatially resolved rotation curves provides the tightest TF relation when compared with other rotational velocity measurements from both spatially resolved and unresolved observations \citep{F_Lelli2019}. However, the comparative calibrations performed in this paper used a relatively small sample size and did not perform any bias corrections. Moreover, this paper focused on the impact of the different rotational velocity definitions on the scatter around the Baryonic \TF (BTF) relation, which characterises the correlation between the total baryonic mass of the galaxy ($M_{\rm baryonic} = M_{\rm stellar} + M_{\rm gas})$ and its rotational velocity, rather than just the stellar mass, which measured via the galaxy's luminosity. Samples selected for calibration of the BTF relation generally consist of larger galaxies with smaller fractions of gas and may vary quite significantly from those used to calibrate the TF relation. 

Hence, in this paper, we have provided a new calibration of the TF relation in the W1 ($3.4\mu$m) and W2 ($4.6\mu$m) bands, which aims to address some of the key concerns with previous calibrations of the TF at mid-infrared wavelengths. In particular, we have used the largest sample to date, consisting of 886 galaxies from 31 different clusters, performed a thorough bias corrections for account for the statistical biases in the sample, and have calibrated the TF relation using both isophotal and total magnitudes in order to determine which of these apparent magnitudes measurements provides the tightest fitting TF relation. We have also compared our calibrated TF relation with previous mid-infrared calibrations as well as those calibrated in the I, J, H and K bands, in order to gain insights into aspects of galaxy evolution. 

This paper is organised as follows: in section~\ref{Section: Data} we discuss the selection criteria for the calibration sample, as well as the processing of the new WISE data. The TF fitting procedure and incompleteness bias correction are discussed in  sections~\ref{Sect:Fitting_Proceedure} and \ref{Section: Incompleteness Bias}. In section~\ref{Section: Morph Type} the correction for morphological galaxy type is discussed. Finally, the fully-calibrated TF relation is compared with the calibrated relations in the near-infrared and optical bands in order to investigate the implications for spiral galaxy evolution in section~\ref{Section: Galaxy Evolution}.

\section{WISE Calibration Sample} \label{Section: Data}

The calibration sample that we use in this paper is the same as the 2MASS \TF calibration sample from the all-sky catalogue, which was used by \cite{K_Masters2008} to calibrate the TF relation in the near-infrared J, H and K bands. This initial sample was selected by cross-matching the 2MASS Extended Source Catalogue with both the Cornell HI digital archive \citep{C_Springob2005} and the database of optical rotation curves \citep{B_Catinella2005}. Three cuts were applied to this initial sample:

\begin{enumerate}
    \item A morphological type cut, to remove galaxies that are classified as having irregular morphology by visual inspection.
    
    \item An inclination cut, to remove galaxies with an inclination of less than $25^{\circ}$ from side on. 
    
    \item A "cluster" cut, to remove field galaxies (that are not located within galaxy clusters) or any galaxies further than 2 Abell radii from the cluster. Any galaxies in clusters that contained less than $\sim 5$ galaxies (within the sample) were also removed.  
\end{enumerate}

These three cuts were applied to the calibration sample to minimise both the scatter in the TF relation and the statistical biases in the data. Irregular spiral galaxies were excluded as they do not generally have the same physical properties as the overall population of spiral galaxies and may therefore introduce additional scatter or bias into the calibration of the TF relation. Spiral galaxies with small inclinations were also excluded as these galaxies tend to have very large inclination corrections, which are not accurate. Finally, cluster galaxies were used for two reasons. First, because the diameter of a cluster is negligible compared to its luminosity distance, the luminosity distance to a cluster can be used as the luminosity distance to an individual galaxy within the cluster. This means that the calibration sample could be expanded to include galaxies where this measurement does not exist or has a very large uncertainty, thereby improving the statistical accuracy of the calibration. Second, although cluster samples suffer from a statistical incompleteness bias, there is a well-established and simple correction technique for this \citep{R_Giovanelli1997}. This avoids attempting to correct for the Malmquist bias in samples of individual galaxies, which is difficult. The cluster assignment used for this sample follows the group assignments described in \cite{C_Springbob2007}. The cluster assignments in this catalogue are derived from 3 sources. The majority of the galaxies in the sample come from the SFI++ template sample, for which groups were assigned based on visual inspection of the transverse and radial distributions of the galaxies in redshift space. Additional galaxies in the 2MTF template sample not included in SFI++ sample were then used to supplement the existing 31 clusters from the SFI++ sample using the same inspection process \citep{C_Springbob2007,K_Masters2008}.

We will be using the mid-infrared as a probe of the stellar content of the spiral galaxy sample.  The fundamental assumption is that the $3-5 \mu m$ tail of the Rayleigh-Jeans distribution is an effective tracer of the dominant stellar mass population.
To obtain our calibration sample, we cross-matched the 2MTF calibration sample with the custom-built WISE Extended Source Catalogue \citep[WXSC;][]{T_Jarrett2013, T_Jarrett2019}, to obtain a sample comprising a total of 888 spiral galaxies from 31 clusters, with known luminosity distances, good quality WISE W1 ($3.4\mu m$) and W2 ($4.6 \mu m$) apparent magnitudes, and rotational velocity measurements. This sample is identical to the 2MTF calibration sample. The surface brightness sensitivity of WISE is much greater than that of 2MASS, and hence the stellar mass probe should be better matched to the extended neutral hydrogen distribution, resulting in a lower-scatter relation.

\begin{figure*}
    \centering
    \includegraphics[scale=0.7]{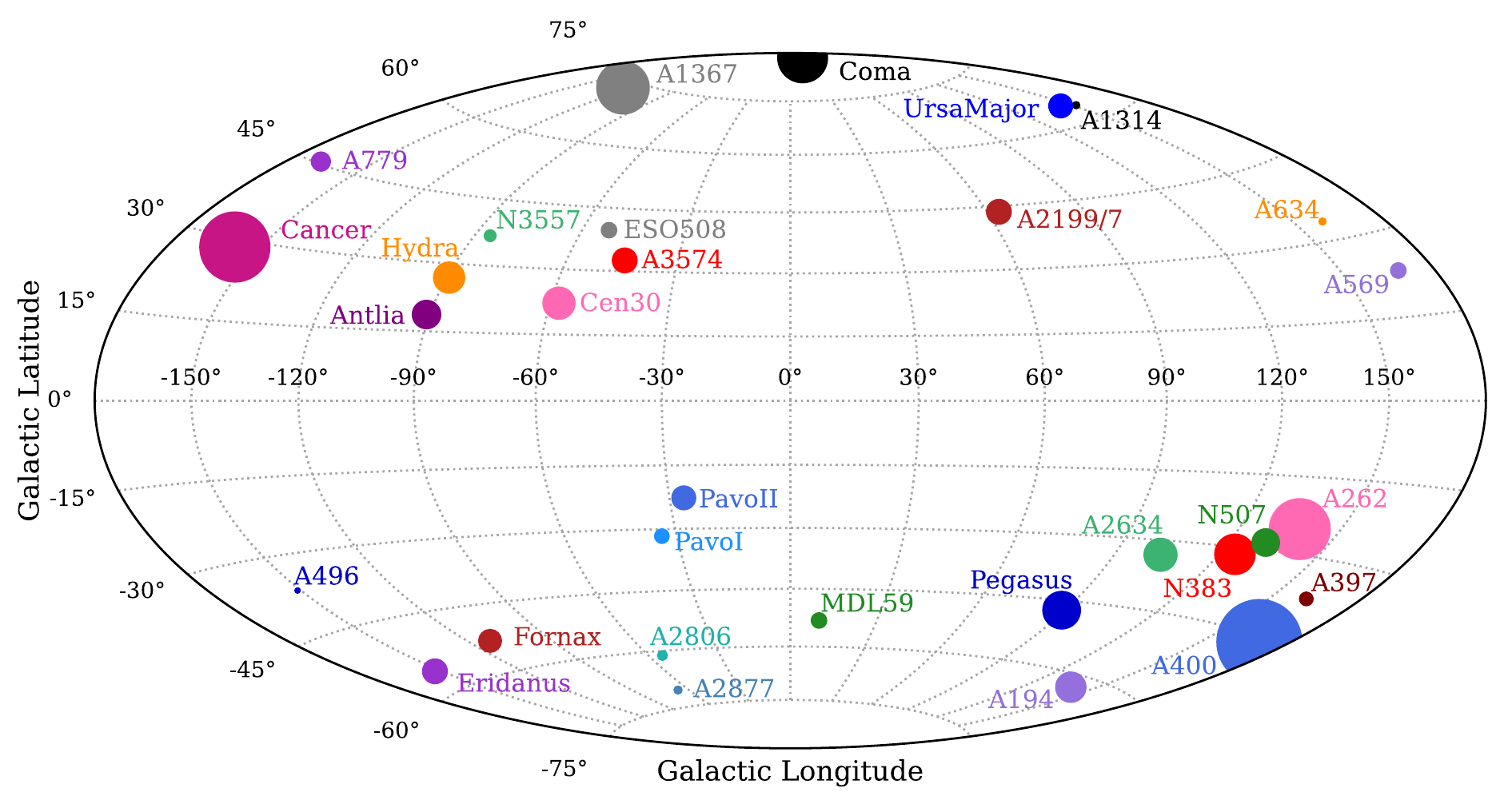}
    \caption[Sky-projection of galaxy clusters in the calibration sample]{The angular positions of the calibration sample on a 2D Aitoff-Hammer projection of the sky. In this plot the radius of the point representing each galaxy cluster is proportional to the number of galaxies from that cluster that are included in the calibration sample.}
    \label{Fig:SampleDist}
\end{figure*}

A plot of the angular positions on a 2D projection of the sky for the galaxy clusters included in the calibration sample can be found in Fig.~\ref{Fig:SampleDist}.

\subsection{Rotational Velocity Measurements}\label{Section:RotationalVelocties}
The rotational velocity measurements were obtained from the 2MASS TF (2MTF) all-sky catalogue \citep{K_Masters2008}, which contains a combination of \hi global profile measurements \citep{C_Springob2005} and optical rotational velocity widths \citep[ORVs;][]{B_Catinella2005}. In the 2MTF catalogue, where good quality \hi linewidths and ORV widths are both available, the \hi linewidth is preferred as the \hi disk typically traces the full extent of the gas layer in spiral galaxies, which comprises a large fraction of the interstellar medium. Moreover, the \hi disk generally extends out to approximately twice the optical size of the galaxy. Thus, it should more effectively trace the rotation curve of galaxies up to large radii and is representative of the majority of the matter in the galaxy \citep{K_Masters2008, M_Haynes2018}. However, a recent paper by \cite{F_Lelli2019}, has shown that the total scatter around the baryonic \TF relation was reduced when the average rotational velocity measured along the flat region of spatially resolved optical rotation curves ($V_f$) was used as the rotational velocity parameter when calibrating the TF relation. Therefore, in section~\ref{Section: Vf}, we also investigate the impact of using $V_f$ as the rotational velocity measurement in the TF relation, by converting the \hi linewidths from the 2MTF catalogue to $V_f$ measurements using the empirical conversion provided by \cite{F_Lelli2019}. 

\subsubsection{ \hi Global Profiles} \label{Sect: HI Profiles}
The observed \hi linewidths ($W_{\rm obs, F50}$) in the 2MTF catalogue were calculated from the \hi global profiles by taking the difference between the points on either side of the \hi profile at which the \hi profile (fitted with a polynomial) is equal to 50\% of the value of $f_p - rms$ for the profile, where $f_p$ is the maximum flux level and $rms$ is the r.m.s scatter of the observed data measured in the signal-free region of the profile \citep{C_Springob2005}. These linewidths were then corrected for instrumental and noise effects ($\Delta_s$), relativistic broadening due to the galaxies' cosmological redshifts ($z$), the impact of turbulent motion ($\Delta_t$), and the disk inclination ($i$). The fully corrected  \hi linewidth ($W_{F50}$) is given by, \citep{C_Springob2005, C_Springbob2007}
\begin{equation}
    W_{F50} = \frac{1}{\sin(i)}\bigg(\frac{W_{\rm obs, F50} - \Delta_s}{1 + z} - \Delta_{\rm t}\bigg).
\end{equation}

Throughout this paper, the fully corrected linewidth $W_{F50}$ will be denoted by $W$. 

A full description of the processing of the raw \hi global profiles and the measurement of the \hi linewidths can be found in \cite{C_Springob2005}. 

\subsubsection{Optical Rotational Velocity Widths}\label{Section:ORCs}
The optical rotational velocity widths are obtained from the optical rotation curves of spiral galaxies, which are plots of the rotational velocity of the disk measured using optical light (generally the H$\alpha$ spectral lines) as a function of radius. The raw rotational velocity widths in the 2MTF catalogue were processed by \cite{B_Catinella2005}, and are corrected for cosmological broadening and inclination following the correction technique for \hi linewidths described in \cite{C_Springob2005}. The ORV width measurements are also converted to units equivalent to the \hi linewidths.

As discussed by \cite{B_Catinella2007}, there is a systematic variation between the rotational velocity widths measured using ORVs and \hi global profiles which, if not corrected for, may introduce a bias into the TF calibration sample. This difference is dependent on the radial extent of the H$\alpha$ emissions ($r_{\rm max}$) as compared to the optical extent of the galaxy ($R_{\rm opt}$), and on the slope of the ORV at the optical radius where the velocity is measured. There are two equations used to correct the ORV widths, which apply to flat and rising (slope greater than 0.5 $\rm km s^{-1} arcsec^{-1}$) ORVs respectively \citep{B_Catinella2007}. These corrections are given by, 
\begin{equation}
\begin{split}
    W_{\rm HI}/W_{\rm ORC} = 1.075 - 0.013 \times \big(r_{\rm max}/R_{\rm opt}\big) \hspace{1cm} &\text{(Flat)},\\
    W_{\rm HI}/W_{\rm ORC} = 0.899 + 0.188 \times \big(r_{\rm max}/R_{\rm opt}\big) \hspace{1cm} &\text{(Rising).}
\end{split}
\end{equation}

\subsection{WISE Photometry}\label{Section:Photometry}
The WISE imaging and source characterization is from the WISE Extended Source Catalogue \citep[WXSC;][]{T_Jarrett2013, T_Jarrett2019}, in which a custom-built deep mosaics are constructed from all of the available WISE imaging of the main survey and the subsequent NEOWISE survey \citep{T_Jarrett2012}. The mosaics have super-sampled 1 arcsec pixels across a 6 arcsec FWHM beam. Source characterisation is then carried out using a pipeline processor, including background and foreground source cleaning, astrometry, size, orientation and photometry measurements, radial profile and surface brightness modeling to assess total emission extent and aggregate flux. Physical parameters, including the host stellar mass and star formation rate are estimated from luminosities that are derived from the k-corrected (redshift) photometry, where the short wavelength bands of WISE are sensitive to the host stellar component and the longer wavelengths the star formation activity \citep{T_Jarrett2013}.

The TF relation is an empirical relation that occurs because both the luminosity and rotational velocity of a galaxy is correlated with its total mass. Therefore, if we were to use the isophotal fluxes, we would be excluding the luminous parts towards the edges of the galaxies in our sample. To account for this, the total integrated (or extrapolated) flux of the galaxy is calculated using a process described by \citep{T_Jarrett2019}. This total integrated flux is generally 5--10 \% larger than the isophotal flux. However, as there has been some debate as to whether this process introduces additional artificial scatter into the TF relation, in section~\ref{Section: Mag_Comparison}, we present a comparison of the TF relations in the W1 and W2 bands calibrated using both isophotal and the total apparent magnitudes, to determine which measurement results in a smaller scatter around the TF relation. 

From this comparison we found that using the isphotal magnitudes did not significantly reduce the scatter around the TF relations. Therefore, because the total flux is more likely to fully characterise the intrinsic luminosity of each galaxy, the calibration of the TF relation described in section~\ref{Section: Incompleteness Bias} uses the total extrapolated fluxes rather than the integrated isophotal fluxes. 

A K-correction was then applied to each of the total flux measurements in order to account for the impact of the cosmological expansion of the universe on the observed fluxes. The  K-correction was calculated in \cite{T_Jarrett2019}, by measuring the WISE filter response (for each WISE bandpass) to synthetic spectral energy distributions at various redshifts for different morphological types. The resulting K-correction, denoted by $K^{\lambda}(z, \rm T)$ is a function of both redshift ($z$) and morphological type (T) \citep{T_Jarrett2017}. The K-correction is then applied to the measured total flux $(f^\lambda_{\rm obs})$ (measured in mJy) using the expression 
\begin{equation}
    f^{\lambda}_{K} = f^{\lambda}_{\rm obs} K^{\lambda}(z, \rm T), 
\end{equation}
where $f^{\lambda}_K$ is the total K-corrected flux (also measured in mJy), and the superscript $\lambda$ refers to the fact that a different correction is needed for each of the wavelengths corresponding to the W1 and W2 bands.
 
The total K-corrected flux for each galaxy was then converted to an apparent magnitude (measured using the Vega magnitude system) via the relation 
\begin{equation}
    m^{\lambda}_{K} = m^{\lambda}_{\rm zero} - 2.5 \log_{10}(F^{\lambda}_{K})
\end{equation}
where $m^{\lambda}_{\rm zero}$ is the instrumental zero-point magnitude derived in \cite{T_Jarrett2011}, and $F^{\lambda}_K$ is the K-corrected flux converted from mJy to digital numbers (DN). The instrumental zero-point magnitudes for the W1 and W2 bands are $m^{3.4 \mu m}_{\rm zero} = 20.5 \text{ mag}$ and $m^{4.6 \mu m}_{\rm zero} = 19.5 \text{ mag}$ respectively \citep{T_Jarrett2017}. 
 
The uncertainties in the corrected magnitudes were calculated via the expression 
\begin{equation}
    \Delta m^{\lambda}_K = \Bigg[ \Big(\Delta m_{\rm zero}^{\lambda}\Big) ^ 2 + 1.179 \bigg(\frac{\Delta f^{\lambda}_K}{f^{\lambda}_K}\bigg)^2\Bigg]^{1/2}, 
\end{equation}
where $\Delta m_{\rm zero}^{\lambda}$ is the uncertainty in the instrumental zero point magnitude, $\Delta f^{\lambda}_K$ is the uncertainty in the corrected flux and the factor 1.179 comes from the term $(2.5/ \ln(10))^2$ used in the error propagation \citep{WISE_UserGuide_2012}.

The K-corrected apparent magnitudes $(m^{\lambda}_K)$ were then adjusted to account for the effects of galactic extinction according to the expression 
\begin{equation}
    m^{\lambda}_{K, g} = m^{\lambda}_K - A^{\lambda}_{g}
\end{equation}
where $A^{\lambda}_g$ is the galactic extinction correction, which adjusts the observed apparent magnitudes to account for the effects of dust in the disc of the Milky Way. $A^{\lambda}_{g}$ is given by $A^{\lambda}_{g} = A^{\lambda} E(B - V)$, where $A^{\lambda}$ is the wavelength-dependent scaling factor equal to $A^{3.4 \mu m} = 0.189$ and $A^{4.6 \mu m} = 0.146$ in the W1 and W2 bands respectively \citep{E_Schlafly2011}. The term $E(B - V)$ is a reddening coefficient (measured in vega mags), which describes the reddening of an observed galaxy due to the absorption of short wavelength light by the dust in the Milky Way. The reddening coefficient is determined based on the position of each galaxy in the sky (relative to the plane of the Milky Way), using the 100$\mu$m \textit{Galactic Dust Extinction Map} recalibrated by \cite{E_Schlafly2011}. 

No internal extinction correction was performed, as previous investigation has indicated that the impact of internal extinction is negligible to the measured apparent magnitude of a galaxy in the W1 and W2 bands \citep{Sakai:2000}. 

\subsection{Outliers}
To identify any potential outliers in the calibration sample, the galaxies were separated into three sub-samples, using the morphological type assignments from visual inspection. Each of these sub-samples was then fit with a preliminary TF relation using the bivariate fitting procedure described in section~\ref{Sect:Fitting_Proceedure}. A sigma-clip was then performed on each of the sub-samples at the 3-sigma level, and the raw WISE images of the galaxies that lay outside of this clip were visually inspected to identify potential obscurations or distortions that may have impacted the photometry. Using this method the galaxies \textit{UGC00433} and \textit{2MASXJ12163954+4605147} were removed from the sample. The other galaxies identified using the sigma clip did not have clearly compromised photometry and were therefore not removed, as one of the aims of this paper is to characterise the intrinsic scatter around the TF relation in the mid-infrared bands.

\section{Bivariate Fitting Technique} \label{Sect:Fitting_Proceedure}
Calibration of the TF relation requires multiple corrections for the statistical and physical biases present in observational measurements, which rely on finding estimates of the fitted TF relation and its intrinsic scatter. Therefore, it is important to establish a fitting procedure which can be consistently applied throughout the calibration process.

The TF relation arises because the total mass of a spiral galaxy is the predominant factor that determines both its luminosity and rotational velocity \citep{M_Aaronson1983}. This implies that the luminosity does not depend directly on the rotational velocity, and vice versa. Therefore, in this paper, the TF relation is characterised using a bivariate fit, which treats neither the absolute magnitude (M) nor the rotational velocity ($\log_{10}(W)$), as a dependent variable. 

Specifically, the TF relation is parameterised as 
\begin{equation} \label{Eqn:TFR_BasicForm}
    M = a - b \log_{10}(W), 
\end{equation}

And the slope $b$ and intercept $a$ of the TF relation are determined via minimisation of the $\chi^2$ merit function

\begin{equation}\label{Eqn:MeritFunction}
    \chi^2 = \mathlarger{\mathlarger{\sum}}_0^{N} \bigg[\frac{M_i - (a - b\log_{10}(W_i))}{\sigma_i} \bigg]^2,
\end{equation}
where $\sigma_i$ is the total uncertainty associated with each galaxy $i$, given by 
\begin{equation}
    \sigma_i = \sqrt{(\Delta M_i)^2 + \left(\abs{b} \cdot \Delta \log_{10}(W_i)\right)^2 + \xi_{\rm int, i}^2}.
\end{equation}
In this expression, $\xi_{\rm int}$ is the intrinsic scatter around the TF relation, which arises due to the inherent variability of the luminosities and rotational velocities of spiral galaxies of a given mass. $\Delta M_i$ is the uncertainty in the absolute magnitude of the $i^{\rm th}$ galaxy and $\Delta \log_{10}(W_i)$ is the uncertainty in the rotational velocity parameter derived from the measurement uncertainty in the rotational linewidth $\Delta W_i$. 

The intrinsic scatter around the TF relation is not a quantity that can be directly observed. Instead, it is generally calculated from the total scatter around the fitted TF relation ($\sigma$). However, the process of fitting the TF relation to any sample of data in order to obtain an accurate estimate of the total scatter around the relation requires knowledge of the intrinsic scatter contribution. Thus, to resolve this cyclical issue, we adopted an iterative fitting approach.

The first step in this approach is to perform a preliminary fit of the TF relation by minimising Eqn.~\eqref{Eqn:MeritFunction}, assuming that there is no intrinsic scatter around the relation (i.e. $\xi_{\rm int} = 0$). This preliminary fit is then used to calculate the residuals around the TF relation. A second order polynomial ($p_\sigma$) was then fit to a plot of the residuals against $\log(W)$ in order to obtain a preliminary estimate of the total scatter around the TF relation $(\sigma)$ as a function of $\log(W)$. Two additional polynomials ($p_{\Delta M}$ and $p_{\Delta \log(W)}$) were then fit to scatter plots of the uncertainties on the absolute magnitudes $(\Delta M)$ and the uncertainties in the $\log(W)$ measurements, multiplied by the slope of the TF relation from the preliminary fit $\left(\abs{b_{\rm prelim}} \cdot \Delta \log(W)\right)$, respectively.

These polynomials were then combined according to the expression 
\begin{equation}
    p_{\xi_{\rm int}} = \sqrt{p_\sigma^2 - p_{\Delta M}^2 - p_{\Delta \log(W)}^2},
\end{equation}

in order to obtain a polynomial model for the intrinsic scatter around the preliminary estimate for the TF relation. 

Once the initial estimate for the intrinsic scatter was calculated, the TF relation was refit to the data using the initial estimate for the intrinsic scatter, $\xi_{\rm in} = \xi_{\rm int}\big(\log_{10}(W)\big)$. This new estimate for the TF relation is then used to recalculate the intrinsic scatter. This process, in which each fit of the TF relation assumes an intrinsic scatter equal to the estimated intrinsic scatter from the previous iteration, is then repeated until the fitted slope and intercept of the TF relation converge.

\begin{figure*}
    \centering
    \includegraphics[width=0.9\textwidth]{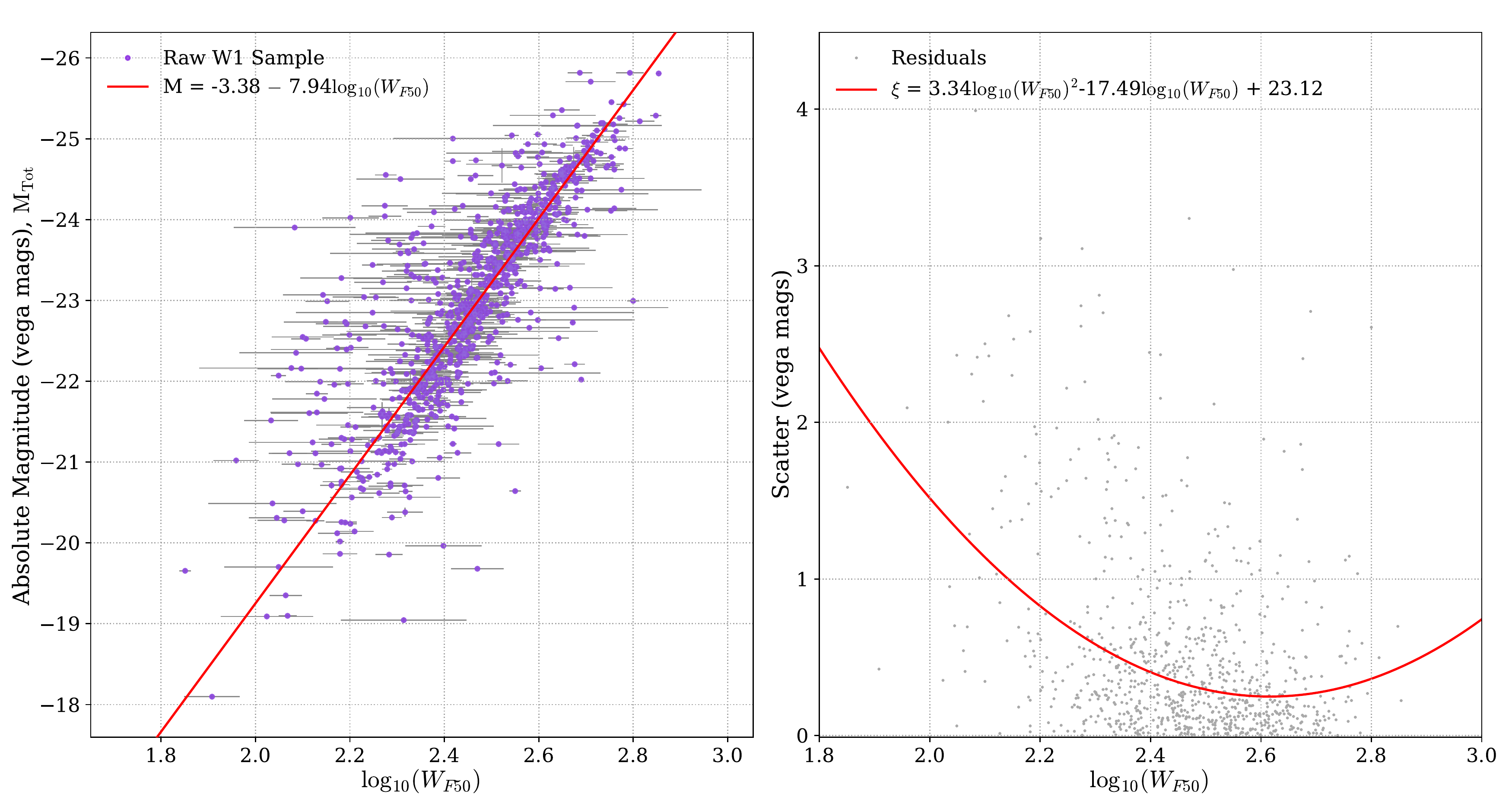}
    \caption[Preliminary bivariate fit to the W1-Band TF calibration sample with the associated intrinsic scatter.]{The left-hand panel shows the preliminary bivariate fit performed on the W1-band calibration sample before and corrections were performed, using the fitting procedure described above. The right-hand panel shows the residuals around the TF relation, with the fitted intrinsic scatter function. At low rotational velocities, the intrinsic scatter function curves sharply upwards due to the impact of the incompleteness bias on the calibration sample. At high rotational velocities, there is also a slight curvature, caused by the combined effects of the morphological dependence of the TF relation, and the lack of Sc galaxies in this region.}
    \label{Fig:Prelim_Fit_W1}
\end{figure*}

Fig.~\ref{Fig:Prelim_Fit_W1} shows the preliminary bivariate fit performed on the W1-band calibration sample, with the associated intrinsic scatter function plotted over the top of the residuals around the line of best fit. In panel 2 of this figure, the intrinsic scatter function is strongly curved at low rotational velocities, due to the impact of the incompleteness bias on the calibration sample. In addition, there is also a slight curvature at high rotational velocities caused by the combined effects of the morphological dependence of the TF relation, and the lack of Sc galaxies in this region.

\section{Cluster Population Incompleteness Bias Correction}\label{Section: Incompleteness Bias}
The selection of only galaxies from clusters in our calibration sample gives rise to a statistical bias known as the \textit{Cluster Population Incompleteness Bias} \citep{P_Teerikorpi1987}, for which the standard bias correction technique was developed by \cite{R_Giovanelli1997}. 

\subsection{Nature of the Incompleteness Bias} \label{Section:Bias_Nature}
Bright galaxies are generally easier to observe than dim galaxies, meaning that a sample of galaxies from a single cluster will generally contain a larger fraction of the bright galaxies in the cluster than the dim ones \citep{R_Giovanelli1997}. As a consequence of this, the distribution of absolute magnitudes in the cluster sample will not accurately reflect the \textit{intrinsic} distribution of absolute magnitudes of the cluster population. At bright magnitudes, the distribution of absolute magnitudes of a cluster sample is expected to closely follow the intrinsic distribution, which is described by the luminosity function. However, as the brightness of a galaxy decreases, it becomes less likely to be observed. Hence, the distribution of magnitudes in a cluster sample is expected to progressively drop away from the luminosity function at dimmer absolute magnitudes. This means that the cluster sample will be increasingly \textit{incomplete} at dim absolute magnitudes. 

When used to calibrate the TF relation, the incompleteness of cluster samples can have a number of significant effects on the fitted slope, intercept, and measured scatter around the TF relation. 

\begin{figure*}
    \centering
    \includegraphics[width=0.9\textwidth]{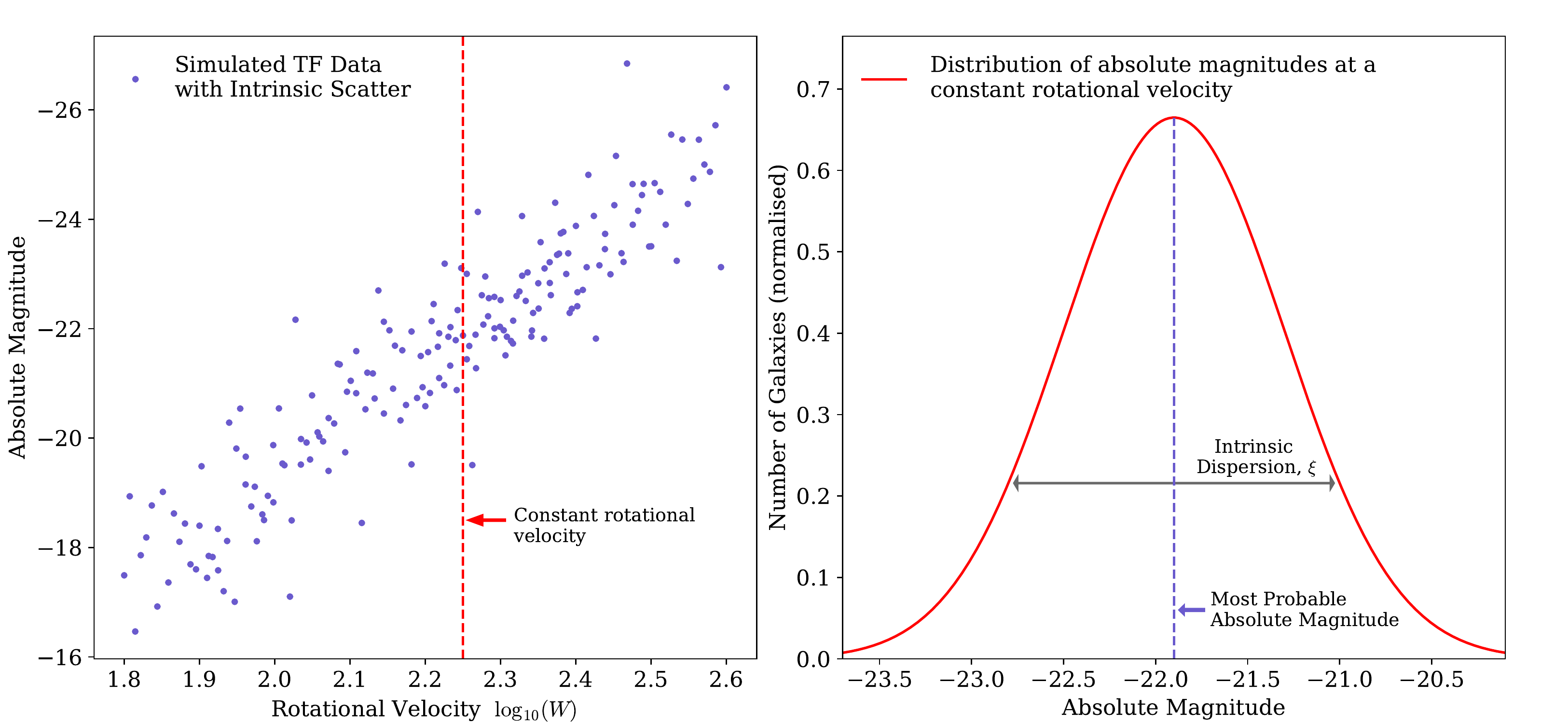}
    \caption[Intrinsic scatter around the TF relation]{This figure demonstrates the intrinsic scatter around the TF relation for a simulated example cluster. The left-hand panel shows the TF relation for an example cluster with a grey dashed line indicating which slice we are considering. The right-hand panel shows the intrinsic magnitude distribution at this rotational velocity.}
    \label{Fig:Magnitude_Dispersion}
\end{figure*}

To illustrate this, we can use a Gaussian distribution to approximate the expected distribution of absolute magnitudes at a given rotational velocity (which leads to the intrinsic scatter around the TF relation, $N_{\rm intrinsic}$). Figure~\ref{Fig:Magnitude_Dispersion} shows such an example of a simulated TF population, along with the corresponding distribution of absolute magnitudes at a vertical slice cut at $\log_{10}(W)$. A full TF sample can be considered as a collection of Gaussian distributions at each rotational velocity in the sample. An unbiased TF relation should pass through the expectation values of each of these distributions at each rotational velocity, and the intrinsic scatter of the TF relation should be equal to the average of the standard deviations for each absolute magnitude distribution.

Because observations of cluster samples are increasingly incomplete at dimmer absolute magnitudes (corresponding to smaller rotational velocities), the observed distribution of absolute magnitudes at low rotational velocities becomes skewed towards brighter absolute magnitudes due to the `loss' of the dim galaxies. Mathematically, the observed distribution of absolute magnitudes is given by 
\begin{equation}
    N_{\rm obs}(M) = c(M) \cdot N_{\rm intrinsic}(M),
\end{equation}
where $c(M)$ is a completeness function which ranges from 1 to 0 and describes the fraction of the true population that has been observed in the sample. 

\begin{figure*}
    \centering
    \includegraphics[width=0.9\textwidth]{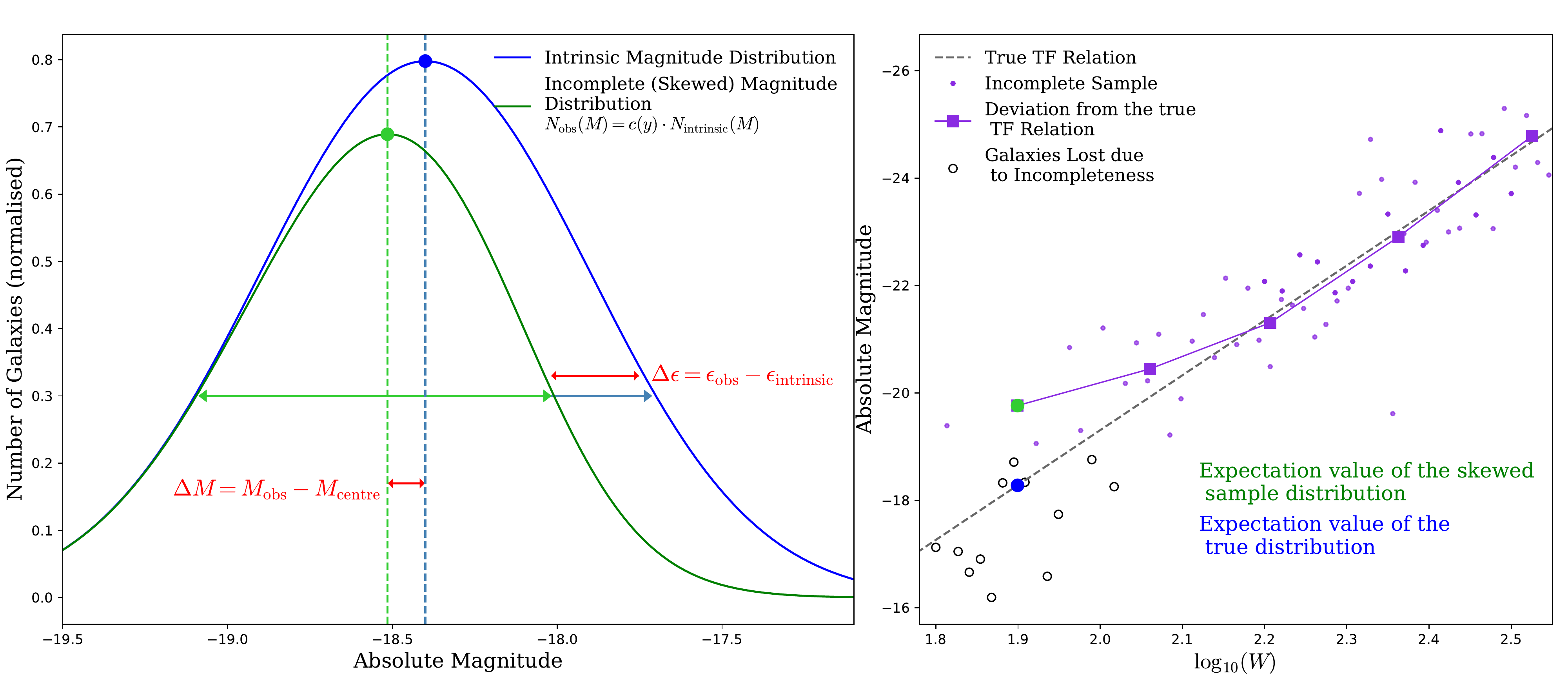}
    \caption[Impact of the incompleteness bias on the TF relation]{This figure shows the impact of the incompleteness bias on the TF relation. The panel on the left shows a comparison between the intrinsic distribution of absolute magnitudes for a given rotational velocity and the skewed distribution of absolute magnitudes from a sample that suffers from incompleteness bias. The shift in the expectation value of the distribution is shown as $\Delta M$ and the decrease in the observed standard deviation is shown as $\Delta \epsilon$. The panel on the right shows an example of an incomplete cluster sample in purple. The large purple squares show the deviation in the moving average from the true TF relation (shown as a dashed line). The galaxies that have been "lost" due to incompleteness bias are shown as open circles. The blue and green points in both figures show the deviation in the expectation value of the sample from the expectation value of the intrinsic distribution as a result of the incompleteness}
    \label{Fig:Skewed_Dispersion}
\end{figure*}

The left-hand panel of Fig~\ref{Fig:Skewed_Dispersion}, shows two example distributions. The blue curve represents the intrinsic distribution of absolute magnitudes for a given rotational velocity and the red curve is an example of a skewed distribution observed for an incomplete cluster sample. 

As the observed distribution of absolute magnitudes become more skewed at lower rotational velocities, the expectation values of observed distributions increasingly deviate from the expectation values of the intrinsic distributions. This effect is shown in the right-hand panel of Fig. \ref{Fig:Skewed_Dispersion}. Here, a simulated incomplete cluster sample is shown. The galaxies that are `lost' due to incompleteness are represented by open circles. In this figure, we can see that at low rotational velocities, the binned average absolute magnitude of the incomplete simulated sample diverges significantly from the intrinsic TF relation, which reflects the deviation between the expectation values of the intrinsic and observed distributions.

The second panel of Fig. \ref{Fig:Skewed_Dispersion} also shows that galaxies that are \textit{observed} at low rotational velocities will systematically have brighter absolute magnitudes than the \textit{true} average absolute magnitude at each rotational velocity. This means that the cluster sample will be biased upwards at low rotational velocities. Thus, the TF relation fitted to this sample will also be biased upwards, resulting in a TF relation with a less steep slope, and a brighter intercept.

\subsection{Incompleteness Bias Correction Technique} \label{Section: Correction Technique}
\begin{figure}
    \centering
    \includegraphics[width=0.48\textwidth]{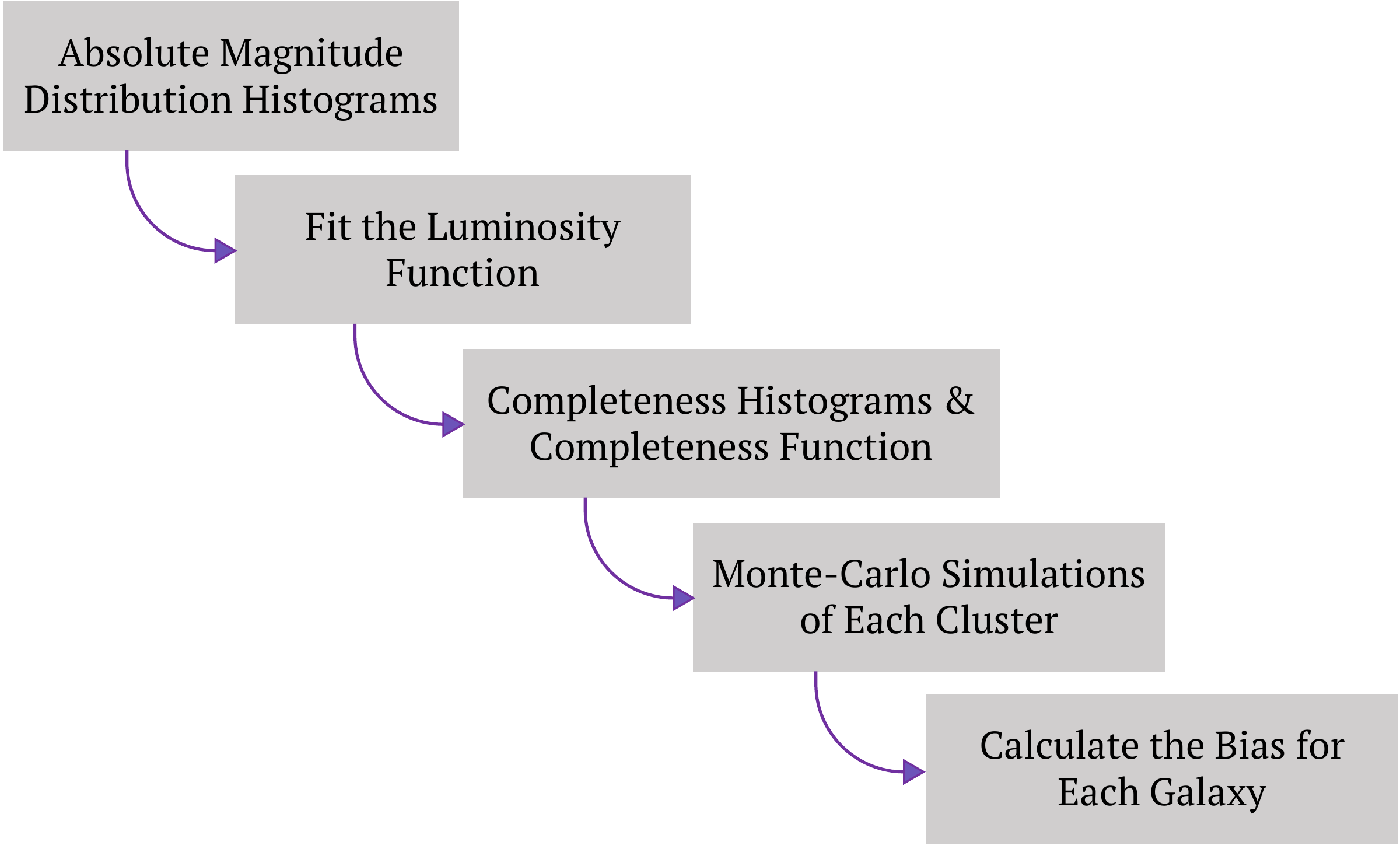}
    \caption[Incompleteness bias correction procedure]{Flow chart illustrating the steps involved in the incompleteness bias correction procedure.}
    \label{Fig:FlowChart_Bias}
\end{figure}

To correct for the incompleteness bias in the calibration sample, we followed the methodology described by \citet{R_Giovanelli1997}. This bias correction technique involves two steps: (1) the characterisation of the incompleteness of each cluster in the calibration sample and (2) the calculation of the bias associated with each galaxy. A flow chart outlining the processes involved in each stage of the calibration can be found in Fig.~\ref{Fig:FlowChart_Bias}. The individual steps of the bias correction were as follows.

For each cluster, a histogram of the distribution of absolute magnitudes was constructed by convolving each absolute magnitude measurement with a Gaussian smoothing kernel, with a standard deviation (width) equal to twice the error associated with the measurement. This process significantly reduced the impact of small number statistics on the constructed absolute magnitude histograms. The probability ($p_{n, i}$) of the measured absolute magnitude of each galaxy being in each bin was calculated by integrating the Gaussian distribution corresponding to each absolute magnitude across each of the bins. The total height of each of these bins was then calculated by summing the probability that each galaxy would have an absolute magnitude measured within that bin 
\begin{equation}
    h_i = \sum_{n=1}^{n=N_g} p_{n, i}.
\end{equation}
Here, $N_g$ is the total number of galaxies in the cluster. The error on the height of each bin was calculated using the expression
\begin{equation}
    \Delta h_i = \sum_{n=1}^{n=N_g} p_{n, i} (1 - p_{n, i}).
\end{equation}

These histograms were then fitted with the luminosity function that describes the intrinsic distribution of absolute magnitudes for that cluster population. The luminosity function for spiral galaxies observed in the W1 and W2 bands is a Schechter function of the form
\begin{equation}
    \Phi(M) = 0.4 \ln(10)  \phi_\star  \Big[10^{0.4(M_\star - M)}\Big]^{\alpha + 1}  e^{-10^{0.4(M_\star - M)}}.
    \label{eq:Schechter}
\end{equation}

In this expression, $\phi_\star$ is the normalisation constant, $M_\star$ is the characteristic absolute magnitude at which the exponential part of the function becomes dominant, and $\alpha$ is the faint-end slope of the function. The parameters $M_\star$ and $\alpha$ are intrinsic properties of the distribution of spiral galaxies when observed in the W1 and W2 bands. As the luminosity functions in the WISE W1 and W2 bands have not been measured, we used the parameterisation of the luminosity functions derived in \citet{X_Dai2009} in the \textit{Spitzer} 3.6$\mu$m and 4.5$\mu$m bands, which closely align with the WISE 3.4$\mu$m and 4.6$\mu$m bands. The parameters that we used were derived using the the parametric maximum-likelihood method and are $M_{\star, 3.6} = -24.27$, $\alpha_{3.6} = -1.40$ and $M_{\star, 4.5} = -24.26$, $\alpha_{4.5} = -1.29$ \citep{X_Dai2009}.

\begin{figure}
    \centering
    \includegraphics[width=0.48\textwidth]{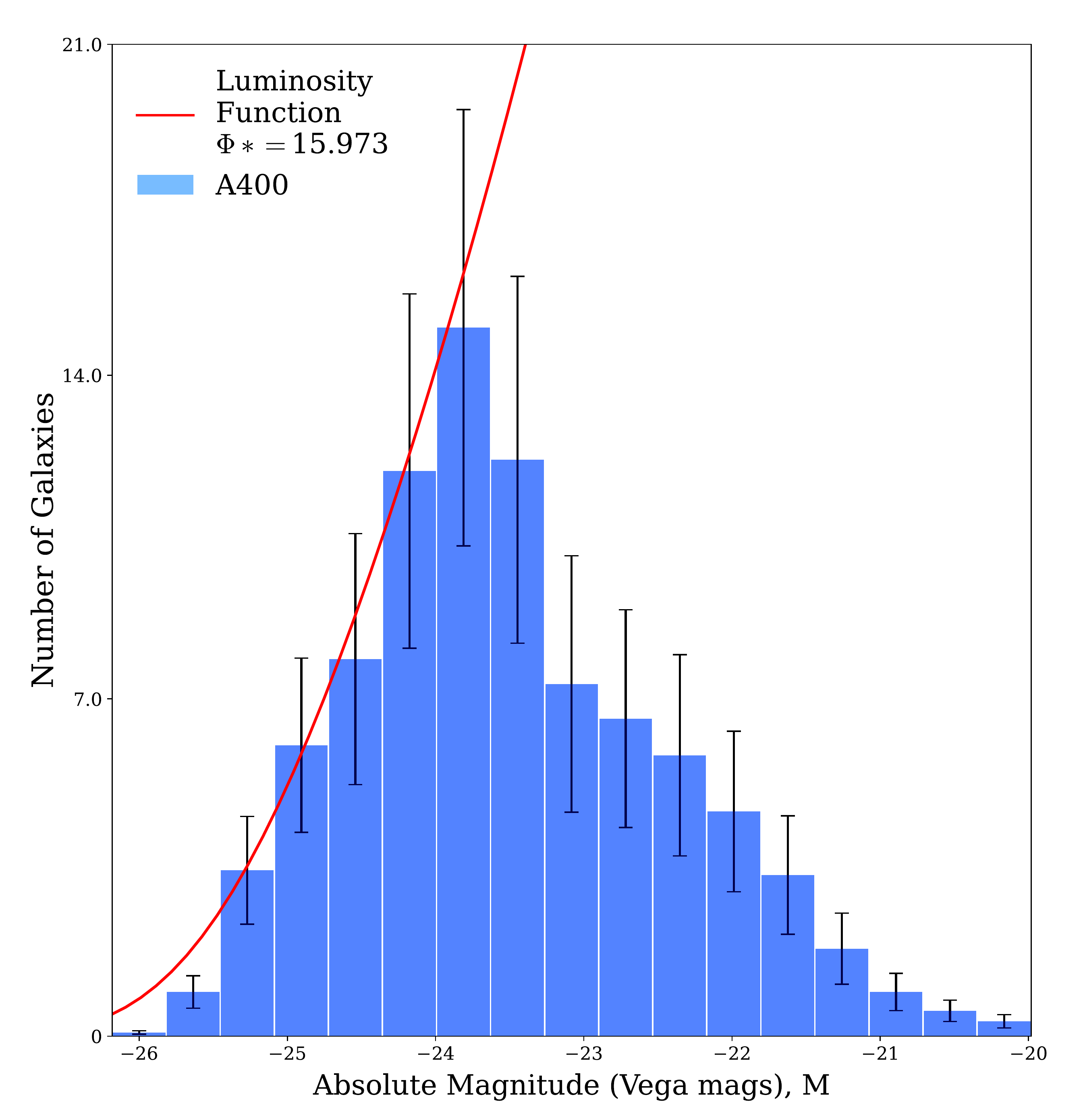}
    \caption[Luminosity histogram and function for the cluster A400]{The distribution of absolute magnitudes for the cluster A400, along with the luminosity function fitted to this histogram. As expected, the luminosity function fits the left side of the magnitude distribution, corresponding to bright galaxies, well. On the right side of the histogram at dim magnitudes, the histogram falls away from the luminosity function, as the cluster becomes more incomplete at dim magnitudes. The first two bins on the left-hand side of the histogram do not closely match the luminosity function because they correspond to the residual probability produced by treating each absolute magnitude as a Gaussian distribution, not to actual galaxy measurements.}
    \label{Fig:AbsMag_Hists}
\end{figure}

The normalisation constant $\phi_\star$ reflects the total number of galaxies in the volume of space that the luminosity function applies to. From preliminary simulations of incomplete cluster samples, we determined that the luminosity function will generally fit the absolute magnitude histogram up to its peak before diverging from the histogram due to incompleteness. Hence, the parameter $\phi_\star$ was determined independently for each cluster by fitting the luminosity function to the bins to the left of the peak of the smoothed absolute magnitude histogram (corresponding to the brighter absolute magnitudes where the sample accurately reflects the cluster population), using $\chi^2$ minimisation with $\phi_\star$ as a free parameter. An example of a absolute magnitude histograms with the fitted luminosity function for the cluster A400 can be found in Fig.~\ref{Fig:AbsMag_Hists}. 

A completeness histogram for each cluster was constructed by dividing the height of each bin of the absolute magnitude histogram by the value of the luminosity function at the centre of the bin.

Each completeness histogram was fitted with a completeness function $c(M)$ of the form
\begin{equation}
    c(M) = \frac{1}{e^{(M-M_F)/\eta}+1}
\end{equation}
where $M_F$ describes the absolute magnitude at which the completeness falls to $50\%$, and $\eta$ describes the steepness with which the completeness function decays.

\begin{figure}
    \centering
    \includegraphics[width=0.48\textwidth]{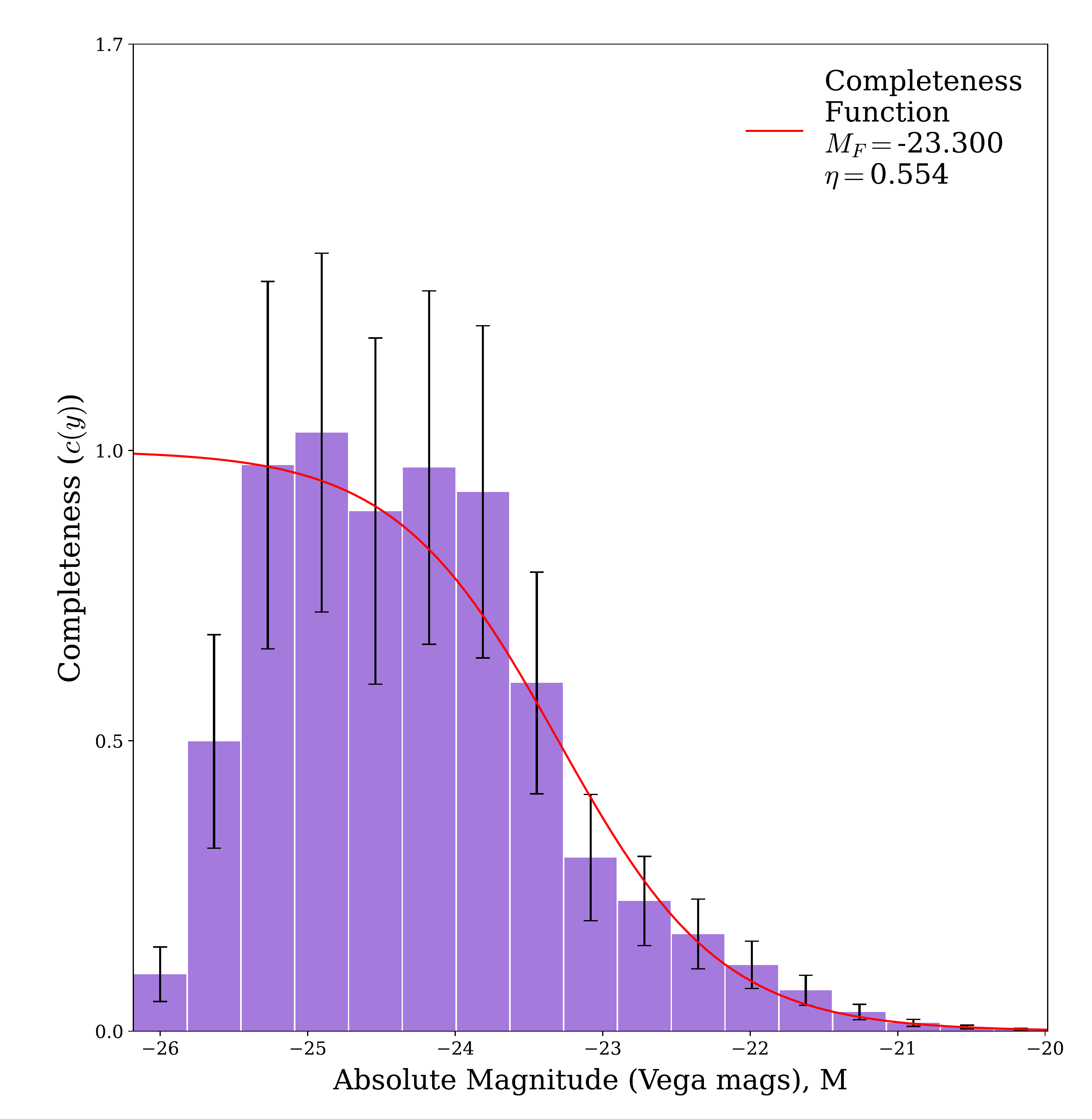}
    \caption[The completeness histogram for the cluster A400]{This figure shows the completeness histogram for the cluster A400, with the fitted completeness function. This function appears to fit the completeness histogram well, especially at dimmer absolute magnitudes, where the cluster becomes more incomplete. As expected, the first two bins on the left-hand side do not match the completeness function and are very small. This does {\em not} indicate that the cluster is incomplete at very bright magnitudes, but is a residual effect of the Gaussian smoothing applied when constructing the absolute magnitude histograms.}
    \label{Fig:Completeness_Hists}
\end{figure}

An analytical function was used when characterising the completeness of each cluster for two reasons: it minimises the effects of small number statistics on the bias correction, and it more accurately characterises the completeness at each individual absolute magnitude as opposed to characterising the incompleteness in discrete steps. An example of a completeness histograms with a fitted completeness function for the cluster A400 can be found in Fig.~\ref{Fig:Completeness_Hists}. 

Once the incompleteness for each cluster was characterised, a Monte Carlo simulation of each cluster was created using a large number of mock samples of the clusters. To create the mocks of each cluster, a preliminary estimate of the slope and intercept of the TF relation (determined using the full calibration sample) was used to predict absolute magnitudes for each galaxy in the cluster using their measured \hi linewidths via the expression 

\begin{equation}
    M_{\rm simulated, i} = a - b \hspace{0.07cm} \log_{10}(W_i) + \xi_{\rm int}.
\end{equation}
In this expression the subscript $i$ refers to the $i^{\rm th}$ galaxy in the cluster. The scatter term in this expression, $\xi_{\rm int}$, was calculated for each galaxy by generating a random number from a Gaussian distribution centred on 0, with a standard deviation given by the functional form of the intrinsic scatter $\xi_{\rm int}(\log_{10}(W))$.

For a given cluster, each mock has the same distribution of \hi linewidths as the observed sample.

Each mock was then passed through a ``completeness filter'' that generated a random number, $n_i$, between 0 and 1 for each galaxy in the cluster. If this random number was greater than the value of the completeness function evaluated at the generated absolute magnitude of the $i^{\rm th}$ galaxy (i.e. if $n_i > c(M_i)$), then the galaxy was removed from the mock. 

This process was repeated 10,000 times for each cluster to produce full Monte Carlo simulations of each cluster with the same statistical incompleteness as the sample.

To calculate the incompleteness bias $\beta_i$ for each galaxy within each cluster, the difference between the simulated absolute magnitude $M_{\text{simulated}, i}$ and the absolute magnitude predicted by the \textit{true} TF relation $M_{\text{predicted}, i}$ was averaged across the mocks
\begin{equation}\label{Eqn:Bias}
    \beta_i = \frac{\mathlarger{\mathlarger{\sum}}_{j=1}^{N_{\text{mocks}, i}} \Big[M_{\text{simulated},i,j} - M_{\text{predicted}, i}\Big]}{N_{\text{mocks}, i}},
\end{equation}
where the subscript $j$ refers to the $j^{\rm th}$ mock simulated absolute magnitude for each $i$ galaxies, and $N_{\text{mocks}, i}$ refers to the total number of simulated absolute magnitudes for a given \hi linewidth (i.e. for a single galaxy) after the mocks have been passed through the completeness filter. The uncertainty associated with the bias was assumed to be 3\% of the magnitude of the bias ($\Delta \beta_i = 0.03 \hspace{0.02cm} |\beta_i|$) in order to account for the statistical uncertainties introduced throughout the process of characterising and correcting for the bias.

The absolute magnitude for each galaxy was then adjusted according to the expression 
\begin{equation}\label{Eqn: AbsMag Bias Corrected}
    M_{\text{corr}, i} = M_{i} - \beta_i,
\end{equation}
where $M_{i}$ is the uncorrected absolute magnitude. The uncertainty in the corrected absolute magnitude is given by
\begin{equation}\label{Eqn: AbsMagErr Bias Corrected}
    \Delta M_{\text{corr}, i} = \sqrt{(\Delta M_i) ^ 2 + (0.03 |\beta_i|)^2}. 
\end{equation}

Any galaxies for which the calculated bias was larger than 5\% of their uncorrected absolute magnitudes were removed from the sample, as biases this large are likely caused by small number statistics at low rotational velocities, which may skew the results. 

Once these galaxies were removed from the sample, the corrected absolute magnitudes and their associated errors were then used to derive a new estimate for the \textit{true} TF relation, to be used in the next iteration of the bias correction procedure. 

\begin{figure*}
    \centering
    \includegraphics[width=0.9\textwidth]{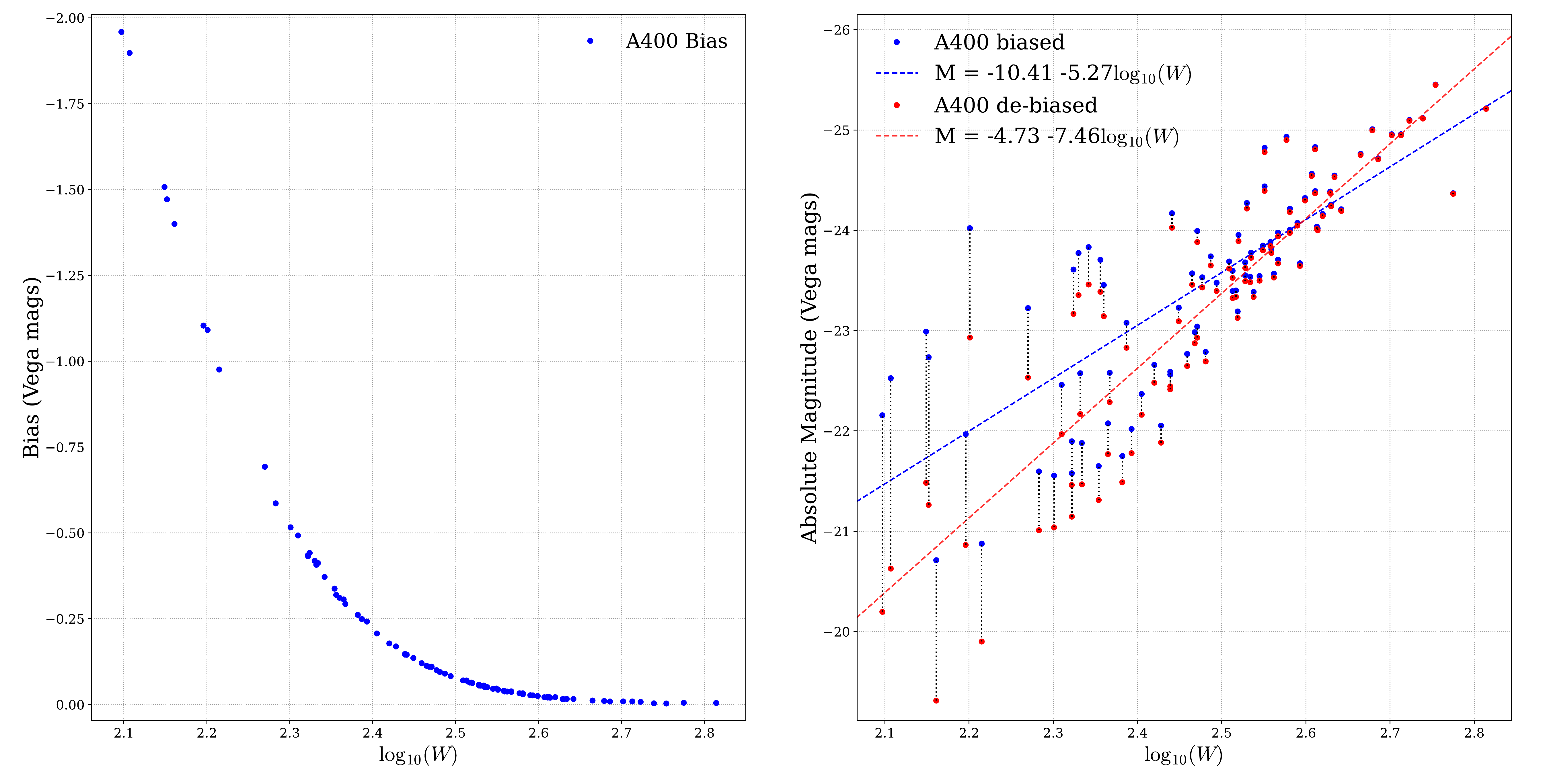}
    \caption[Incompleteness bias for A400 and associated absolute magnitude corrections]{The left-hand panel shows the bias calculated for each of the galaxies in A400, expressed in Vega mags. The magnitude of the bias increases as the rotational velocity decreases, due to the increasing incompleteness of the sample at low rotational velocities (corresponding to dim absolute magnitudes). The right-hand panel shows the original and adjusted absolute magnitude of each of these galaxies. The bias correction has shifted the galaxies at low rotational velocities downwards so that they lie more closely in line with the galaxies at high rotational velocities, which has the effect of steepening the slope fitted to this individual cluster sample.}
    \label{Fig:Biases}
\end{figure*}

A plot of the incompleteness bias for the example cluster A400 can be found in Fig.~\ref{Fig:Biases}.

\subsection{The Full Incompleteness Bias Correction}\label{Section: BiasResults}

\begin{figure}
    \centering
    \includegraphics[width=0.45\textwidth]{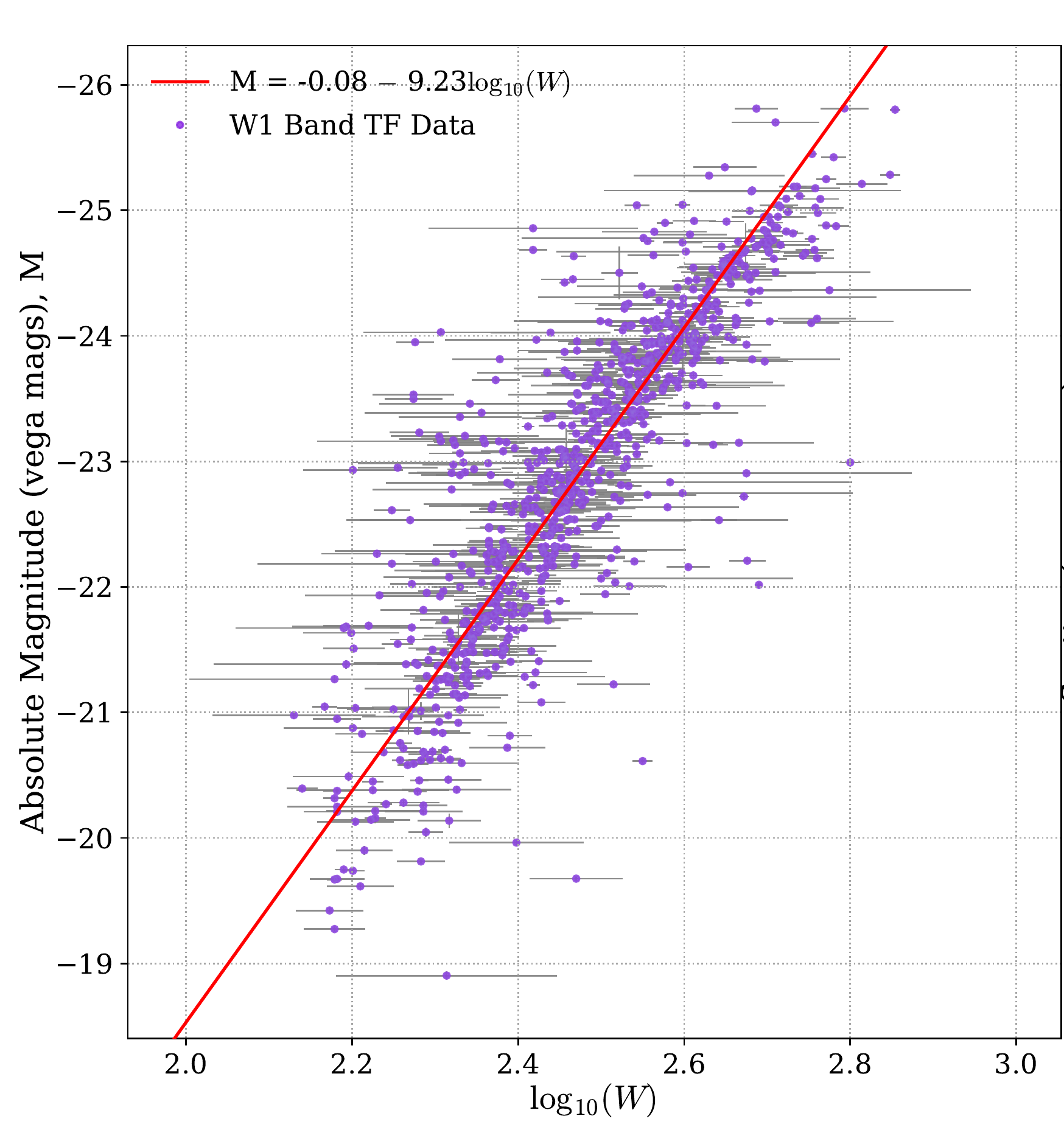}
    \caption[Bias corrected TF W1 band sample ]{Fully bias-corrected TF sample in the W1 band. The bias correction has steepened the slope (as expected), and brought the majority of the low rotational velocity galaxies closer to the x axis so that they are closer to the trend line seen for the bulk of the medium to high rotational velocity galaxies. }
    \label{Fig: Bias_Corrected W1}
\end{figure}

\begin{figure}
    \centering
    \includegraphics[width=0.45\textwidth]{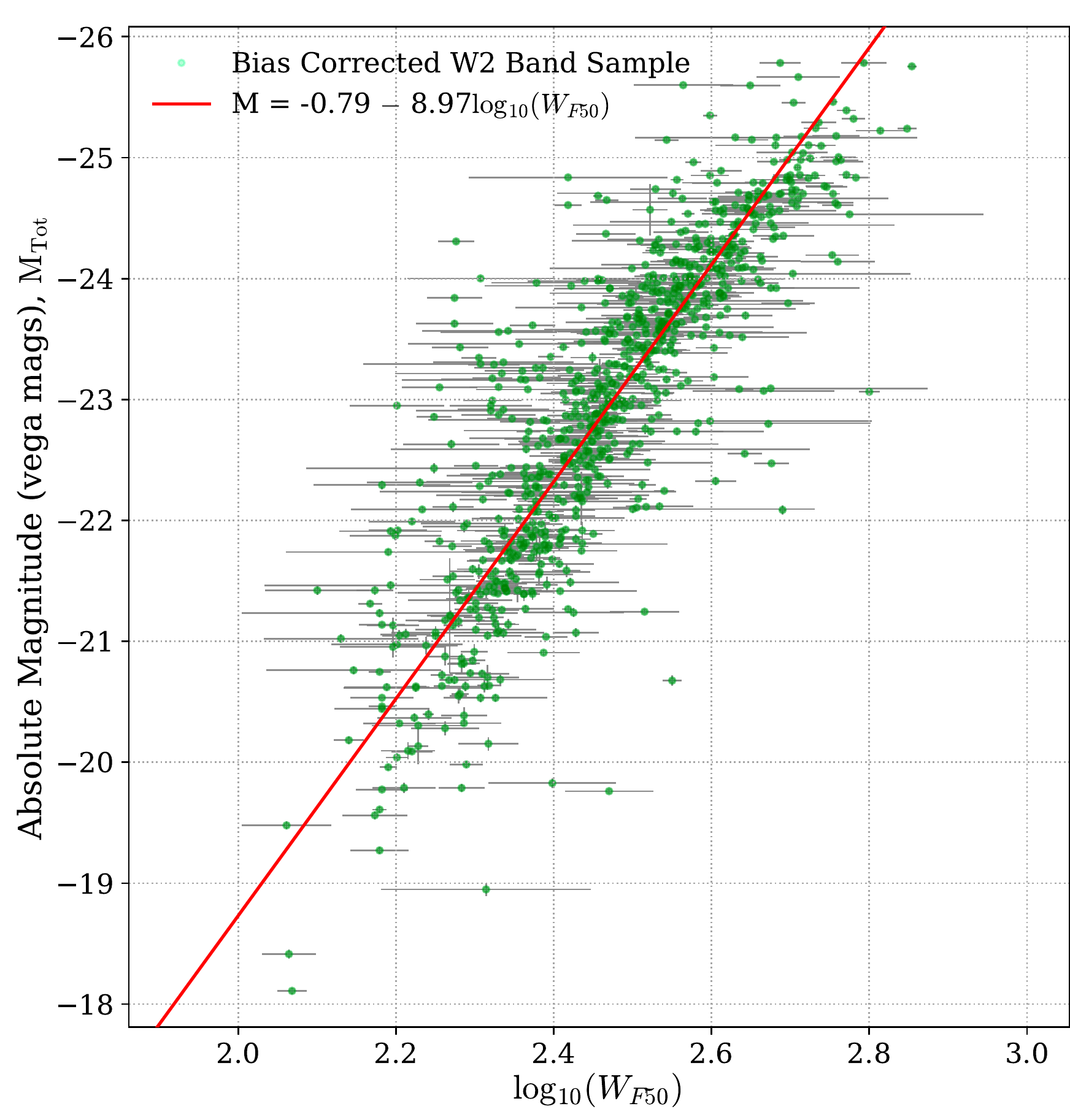}
    \caption[Bias corrected TF W2 band sample ]{As in Fig.~\ref{Fig: Bias_Corrected W1}, this figure shows the fully bias-corrected TF sample in the W2 band compared with the uncorrected sample. The TF relation fit to this corrected sample is not as steep as the relation fit to the W1 band sample}
    \label{Fig: Bias_Corrected W2}
\end{figure}

Figures~\ref{Fig: Bias_Corrected W1} and \ref{Fig: Bias_Corrected W2} show the fully bias-corrected calibration sample in the W1 and W2 bands. In both mid-infrared bands, the bias correction procedure was performed once, rather than multiple times as the steep tail-end of the luminosity function in these bands makes it almost impossible to iterate the bias correction procedure without significantly overestimating the biases on low rotational velocity galaxies. After the bias corrections were performed, the calibration sample for the W1 band contained 848 galaxies, and the calibration sample for the W2 band contained 857 galaxies. 

Overall, the full bias correction changed the slope of the W1 band TF relation by 1.31 Vega mags, and the W2 band slope by 1.12 Vega mags. The TF relations fitted to the bias-corrected calibration sample in the W1 and W2 bands are
\begin{equation}
\begin{split}
    M_{\rm W1} &= (-0.08 \pm 0.45) - (9.23\pm 0.18) \log_{10}(W)\\
    M_{\rm W2} &= (-0.79 \pm 0.45) - (8.97\pm 0.18) \log_{10}(W).
\end{split}
\end{equation}

The total scatter around each of these relations is $\sigma_{W1, \rm debiased} = 0.68 $ and $\sigma_{W2, \rm debiased} = 0.70$, which are reduced from $\sigma_{W1, \rm raw} = 0.77 $ and $\sigma_{W2, \rm raw} = 0.78$. 

The incompleteness bias correction significantly improved the fit of the TF relation in both the W1 and W2 bands. In particular, the scatter around the low rotational velocity end of the relation has been significantly reduced and the best fit TF relations now pass through the majority of the data points in the calibration sample.

\section{Morphological Type Correction}\label{Section: Morph Type}
The TF relation has an intrinsic morphological dependence, which manifests as the downwards curvature at the high rotational velocity end of the TF relation. This is because Sa and Sb type galaxies are generally larger than Sc type galaxies, meaning that there is a larger concentration of Sa and Sb type spiral galaxies at high rotational velocities. However, these galaxies have also been shown to have systematically lower luminosities than Sc type galaxies for any fixed rotational velocity \citep{S_Shen2009}. Hence within a given sample there is generally a higher concentration of these systematically dimmer galaxies at the high rotational velocities \citep{M_Roberts1978}. In addition, Sa and Sb galaxies are weaker \hi emitters and tend to suffer from higher levels of incompleteness than Sc galaxies. These combined effects cause the slopes of the TF relations fitted to the Sa and Sb morphological type sub-samples to be less steep than the slope of the relation fitted to the Sc sub-sample \citep{K_Masters2006,K_Masters2008}.
While its exact cause remains unknown, the morphological dependence of the TF relation is theorised to be the result of the combined effects of different stellar populations between the different types of spiral galaxies and their different disk dynamics \citep{S_Shen2009, C_Tonini2014}.

Correcting for this effect significantly reduces the total scatter around the TF relation, which is important for applications of the TF relation in cosmology.

Because Sc type spiral galaxies are the most common morphological type and less susceptible to small number statistics bias, we corrected the Sa and Sb sub-samples so that their TF relations more closely aligned with the Sc sub-sample's TF relation. 

This correction was performed as follows: the calibration sample corrected for the incompleteness bias was separated according to morphological type and each sub-sample was fitted with an initial TF relation using the bivariate fitting procedure. An additive offset to be applied to each of the Sa $\Delta_{\rm offset, Sa}$ and Sb $\Delta_{\rm offset, Sb}$ galaxies was calculated using the expressions

The morphological dependence is generally assumed to be the result of the large central bulges in Sa (and to a lesser extent Sb) spiral galaxies, which generally have older stellar populations than the disks. When observed at near- and mid-infrared wavelengths, the older stellar populations

\begin{equation}
\begin{split}
    \Delta_{\rm offset, Sa} &= a_{\rm Sa} - a_{\rm Sc}\\
    \Delta_{\rm offset, Sb} &= a_{\rm Sb} - a_{\rm Sc},
\end{split}
\end{equation}
where $a_{\rm Sa}$, $a_{\rm Sb}$ and $a_{\rm Sc}$ are the zero-points of the TF relations fitted to the Sa, Sb and Sc sub-samples respectively. Slope offsets for the Sa and Sb galaxies were calculated using the expressions
\begin{equation}
\begin{split}
\Delta_{\rm slope, Sa}&= b_{\rm Sa} - b_{\rm Sc}\\
\Delta_{\rm slope, Sb} &= b_{\rm Sb} - b_{\rm Sc},
\end{split}
\end{equation}

where $b_{\rm Sa}$, $b_{\rm Sb}$ and $b_{\rm Sc}$ are the slopes of the TF relations fitted to the Sa, Sb and Sc sub-samples.

The complete morphological corrections applied to the Sa and Sb galaxies in the W1 and W2 bands are given in Eqns.~\ref{Eqn:MorphCorrections_W1} and ~\ref{Eqn:MorphCorrections_W2}. 

\begin{equation}\label{Eqn:MorphCorrections_W1}
\begin{split}
    \Delta M_{\rm W1, Sa} &=  7.92 - 3.13 \log_{10}(W)\\
    \Delta M_{\rm W1, Sb} &=  2.38 - 0.99 \log_{10}(W),
\end{split}
\end{equation}

\begin{equation}\label{Eqn:MorphCorrections_W2}
\begin{split}
    \Delta M_{\rm W2, Sa} &=  8.87 - 3.51 \log_{10}(W)\\
    \Delta M_{\rm W2, Sb} &=  3.37 - 1.40 \log_{10}(W).
\end{split}
\end{equation}

\begin{figure}
    \centering
    \includegraphics[width=0.45\textwidth]{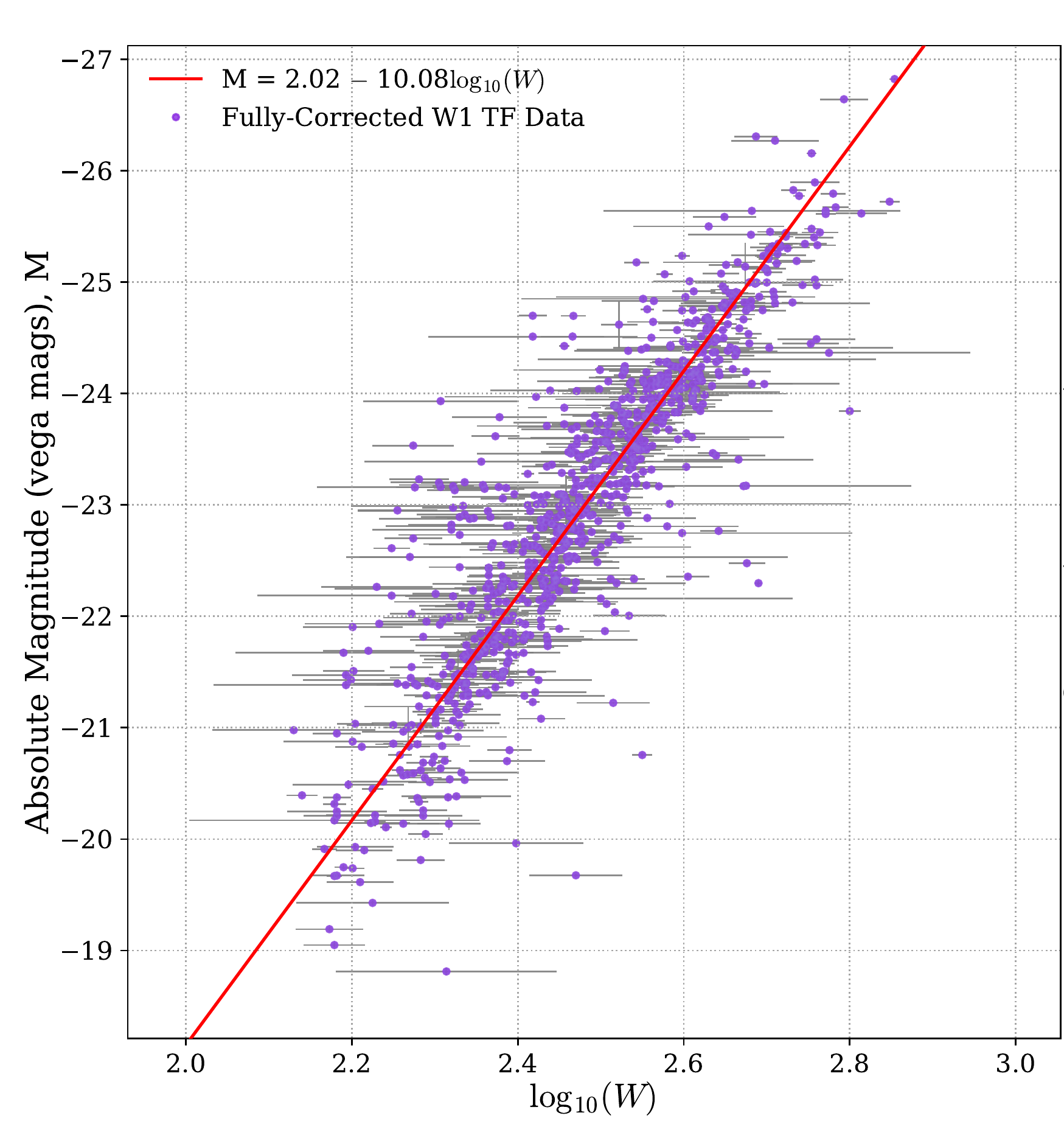}
    \caption[Full morphological correction in the W1 band]{The full TF sample in the W1 band after the incompleteness bias and morphological corrections have also been applied. As expected, the morphological correction has removed the curvature at high rotational velocities, and has increased the slope of the fitted TF relation.}
    \label{Fig:W1_Band Morph Correction}
\end{figure}

\begin{figure}
    \centering
    \includegraphics[width=0.45\textwidth]{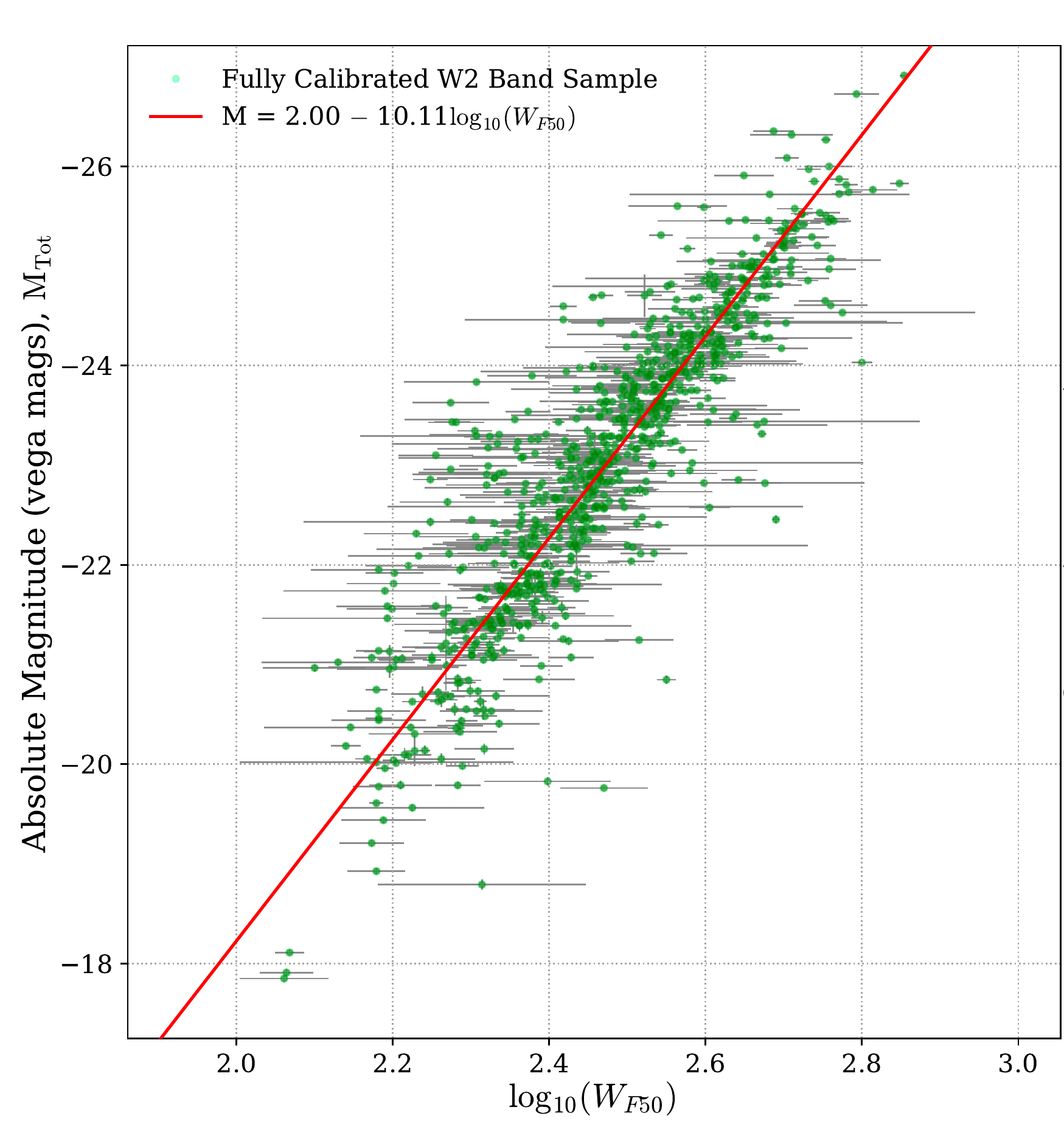}
    \caption[Full morphological correction in the W2 band]{The W2 band after the incompleteness bias correction and the morphological correction. As in the W1 band, the morphological correction has removed the curvature at high rotational velocities, and has increased the slope of the fitted TF relation in the W2 band.}
    \label{Fig:W2_Band Morph Correction}
\end{figure}

Figures.~\ref{Fig:W1_Band Morph Correction} and \ref{Fig:W2_Band Morph Correction} show the fully-calibrated TF relations after both the incompleteness bias and morphological corrections, in the W1 and W2 bands. 

The impact of the morphological correction on the calibration sample is clearly demonstrated in these figures. The fully corrected TF relations are given in Eqn.~\eqref{FULL_TF Relations}.

\begin{equation}\label{FULL_TF Relations}
\begin{split}
    M_{\rm Tot, W1} = (2.02 \pm 0.44) - (10.08 \pm 0.17)\log_{10}(W)\\
    M_{\rm Tot, W2} = (2.00 \pm 0.44) - (10.11 \pm 0.17)\log_{10}(W)
\end{split}
\end{equation}

The total scatter around each of these relations is $\sigma_{W1} = 0.67$ and $\sigma_{W2} = 0.69$, respectively.

% Because of the inherent differences between the three populations of spiral galaxies, it is impossible to fully-characterise the intrinsic TF relation for each population with a single relation. Ideally, the TF relation would be calibrated independently for each morphological type in order to provide accurate distance estimates for spiral galaxies of each type. However, there are a number of factors that prevent accurate calibrations from being performed for morphological type sub-samples. First, separating the TF relation into sub-samples significantly reduces the size of each of these samples, leading to increased statistical errors within the calibration. Second, it is impossible to perform an accurate bias correction on the individual morphological types, as many of the individual clusters would not have enough galaxies of a single morphological type to accurately characterise the cluster's statistical incompleteness. Hence, the best option is to attempt to correct the full TF relation so that it can accurately calculate the distances to the \textit{majority} of spiral galaxies. 

\section{Impact of Rotational velocities derived from Rotation Curves on the total scatter around the TF relation.}\label{Section: Vf}

As discussed in section~\ref{Section: Data}, our calibration sample predominantly contained rotational velocities obtained from unresolved \hi linewidths, supplemented by a small number of measurements from optical rotational curves that were converted to the \hi linewidth scale. However, a recent comparison of the impact of using different definitions of rotational velocity on the baryonic TF (BTF) relation by \cite{F_Lelli2019}, found that of the tightest BTF relation was obtained when using average rotational velocities measured along the flat sections of the rotation curves of the galaxies ($V_f$). The BTF relations obtained using the two definitions of rotational velocity from \hi linewidths, W\textsubscript{P20} and W\textsubscript{M50}, both had shallower slopes and larger total scatters. Therefore, in this paper, we have also tested the impact of using $V_f$ to measure rotational velocities on the scatter around the non-baryonic TF relations in the W1 and W2 bands.

We do not have access to the rotation curves for our full calibration sample. Therefore, we obtained $V_f$ measurements for each of the galaxies in our sample by using the empirical conversion between $V_f$ and $W_{M50c}$ provided by \cite{F_Lelli2019}. This conversion is given by 

\begin{equation}
    \log(V_f) = 0.94\log(\frac{W_{M50c}}{2}) + 0.10.
\end{equation}

$W_{M50c}$ refers to the \hi linewidths measured at 50\% of the mean flux density of the \hi global profile, which has been corrected for instrumental resolution, relativistic broadening, and turbulent motions. While this parameter is not exactly the same as the $W_{F50c}$ measurements that we used in our calibration sample, which is defined in section~\ref{Section: Data}, it is similar enough for us to use the same expression to convert to $V_f$. 

The fully-calibrated TF relations in the W1 and W2 bands using the converted $V_f$ measurements can be found in Figs. ~\ref{Fig:W1_Vrot} and ~\ref{Fig:W2_Vrot} below.

\begin{figure}
    \centering
    \includegraphics[width=0.48\textwidth]{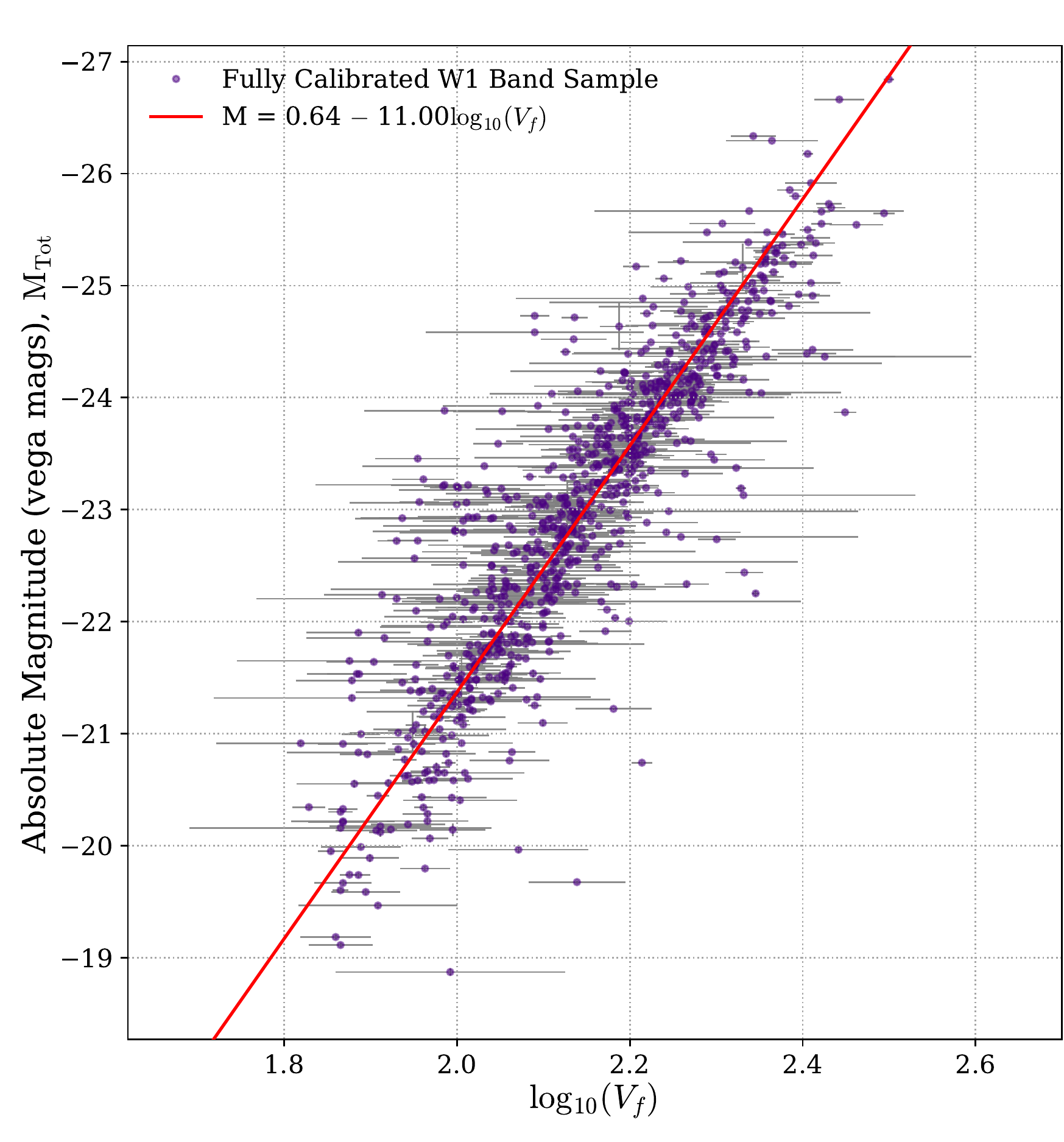}
    \caption[TF relation in the W1 band using $V_f$ to define rotational velocities.]{As expected the transformation to the $V_f$ scale has `compressed' the rotational velocities inwards and shifted them towards smaller values on the horizontal axis. This has significantly steepened the fitted TF relation, which aligns with the results found in \cite{F_Lelli2019}. However, it also appears to have introduced some minor curvature at the high rotational velocity end of the relations that was not present in the calibrations using $W_{F50}$}
    \label{Fig:W1_Vrot}
\end{figure}

\begin{figure}
    \centering
    \includegraphics[width=0.48\textwidth]{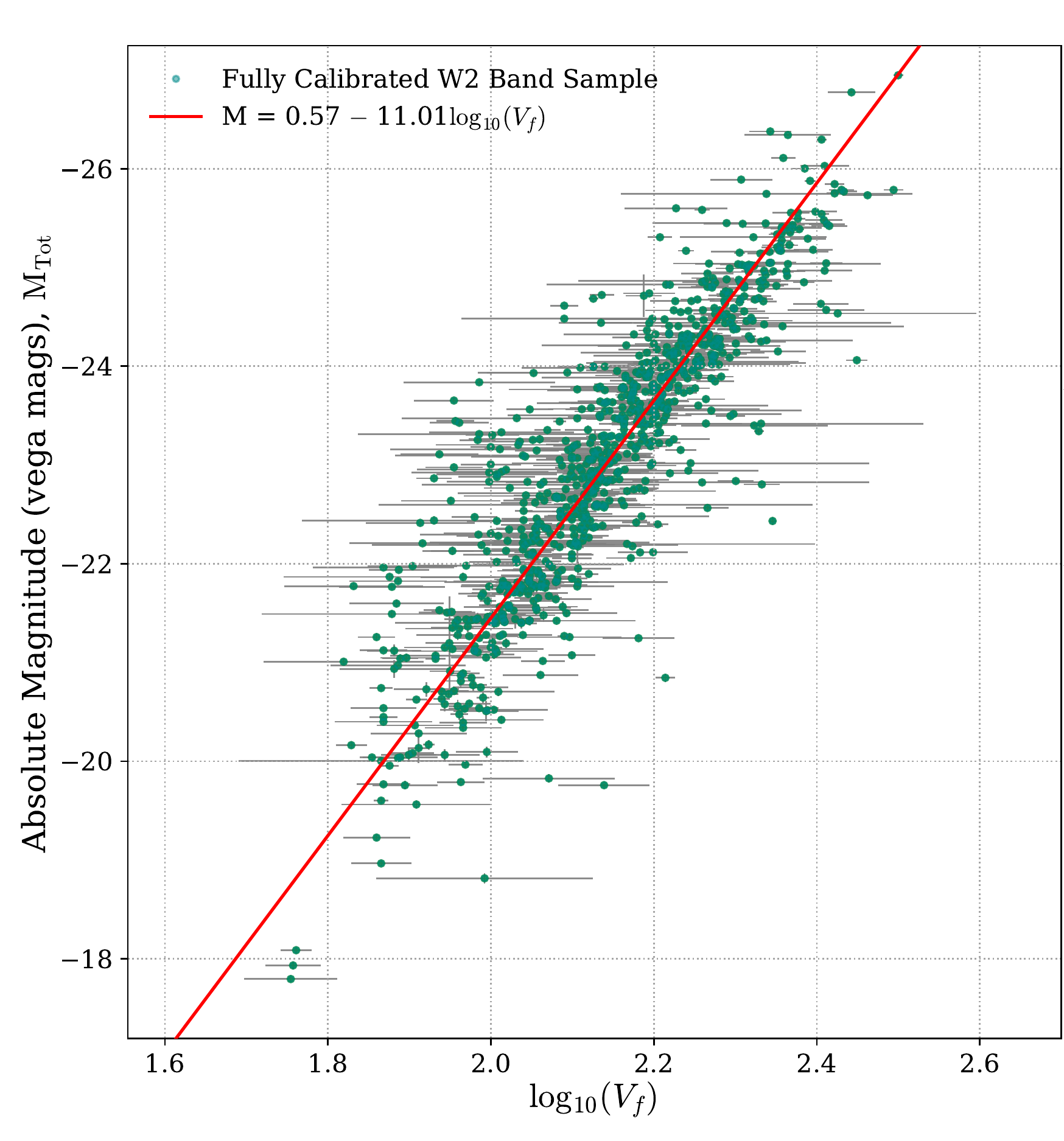}
    \caption[TF relation in the W2 band using $V_f$ to define rotational velocities.]{As with the relation in the W1 band, the TF relation in the W2 band using the converted rotational velocities is steeper than that fit using $W_{F50}$ and has a slight curvature. }
    \label{Fig:W2_Vrot}
\end{figure}

These relations are given by

\begin{equation}
\begin{split}
    M_{\rm Tot, W1} = (2.02 \pm 0.44) - (11.00 \pm 0.20)\log_{10}(V_f)\\
    M_{\rm Tot, W2} = (0.57 \pm 0.43) - (11.01 \pm 0.20)\log_{10}(V_f)
\end{split}
\end{equation}

with total scatters of $\sigma_{W1} = 0.68$ and $\sigma_{W2} = 0.70$ respectively. In line with the results found by \cite{F_Lelli2019}, the slopes of these calibrated TF relations were steeper than those obtained using the $W_{F50c}$ measurements. However, the total scatters associated with these fits were larger by 0.02 and 0.01 in the W1 and W2 bands respectively, which while only marginal, is the opposite of the findings of \cite{F_Lelli2019}. 

The slightly increased total scatter around these relations as compared with those calibrated using $W_{F50}$ to measure rotational velocities may have been caused by a number of potential factors. Firstly, it is possible that the systematic differences between the $W_{F50c}$ and $W_{M50c}$ rotational velocities are large enough that the conversion derived by \cite{F_Lelli2019} cannot be validly applied to our sample, resulting in some additional statistical uncertainty in the calibration. Secondly, because the baryonic TF relation generally does not include as many low surface brightness galaxies as our sample, which are dominated by gas and generally occur at low rotational velocities, it does not exhibit the significant scatter at low rotational velocities that is present in our TF relation.

\section{Impact of Isophotal magnitudes on the total scatter around the TF relation.}\label{Section: Mag_Comparison}
The isophotal fluxes in the W1 and W2 bands are derived by integrating the observed flux for each galaxy within the 1-$\sigma$ sky ($\sim$23 mag arcsec\textsuperscript{-2}) isophotes fit in the W1 band. In contrast, the total extrapolated fluxes are calculated by integrating out a double Séric function fit to the surface brightness profile of the galaxy to a radius of three disk scale lengths beyond the 1-$\sigma$ sky isophotal limit. Both the total extrapolated and isophotal fluxes have been used when calibrating the TF relation in the past, with varied results, particularly in the magnitude of the total scatter around the relation. Hence, in the interest of determining the impact of using these different apparent magnitude measurements on the fitted TF relation and its associated total scatter, we have also run the calibration using the isophotal magnitudes in both the W1 and W2 bands.

\begin{figure}
    \centering
    \includegraphics[width=0.45\textwidth]{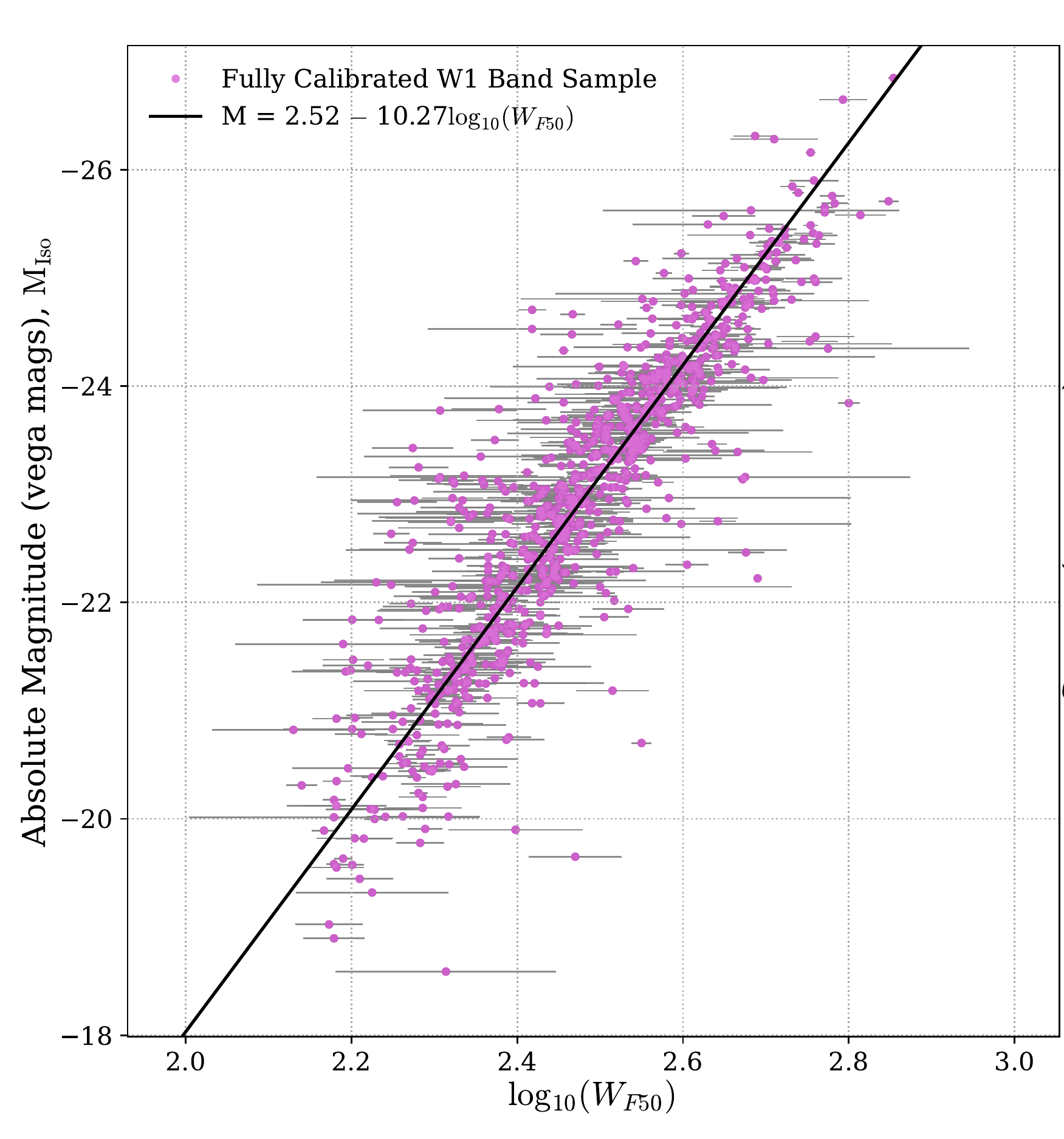}
    \caption[Fully Calibrated TF Relation in the W1 band with Isophotal Magnitudes]{The TF relation in the W1 band using isophotal is slightly steeper than the Relation calibrated using total magnitudes. There also appears to be a very faint downwards curvature at low rotational velocities in this relation that was not present in the fully calibrated sample using total magnitudes. }
    \label{Fig:W1-ISO}
\end{figure}

\begin{figure}
    \centering
    \includegraphics[width=0.45\textwidth]{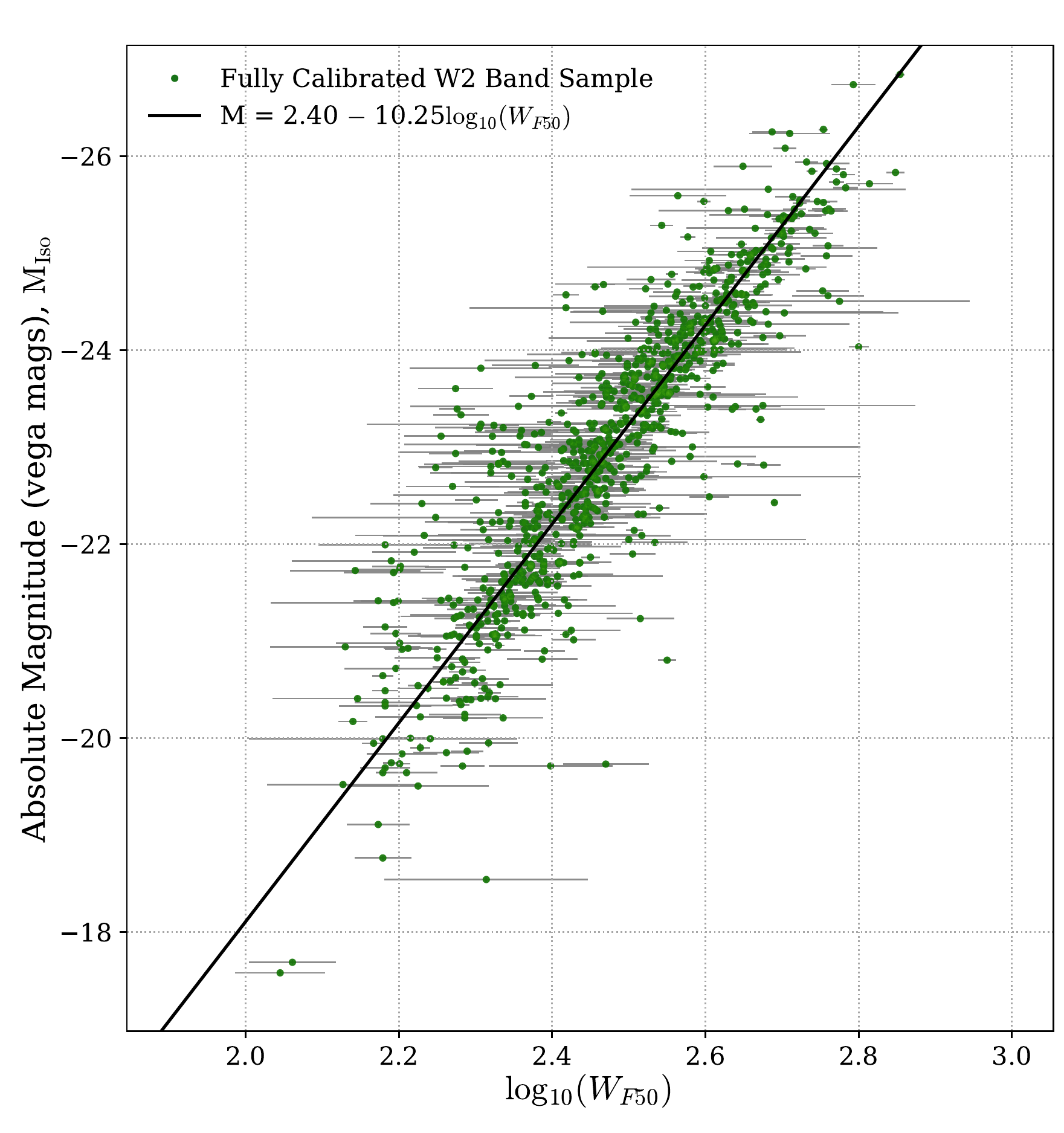}
    \caption[Fully Calibrated TF Relation in the W2 band with Isophotal Magnitudes]{As with the TF relation with isophotal magnitudes in the W1-band, in the W2 band, the fully corrected TF relation is slightly steeper when calibrated using isophotal magnitudes and appears to have a very slight downwards curvature at low rotational velocities. }
    \label{Fig:W2-ISO}
\end{figure}

The calibration of the TF relation using the isophotal magnitudes in both the W1 and W2 bands were performed following the methods described above. Figures~\ref{Fig:W1-ISO} and \ref{Fig:W2-ISO} show the fully corrected samples, with the fitted TF relation using isophotal magnitudes for the W1 and W2 bands. 

The fully corrected TF relations in the W1 and W2 bands are given by 

\begin{equation}\label{FULL_TF Relations_ISO}
\begin{split}
    M_{\rm Iso, W1} = (2.52 \pm 0.45) - (10.27 \pm 0.18)\log_{10}(W)\\
    M_{\rm Iso, W2} = (2.40 \pm 0.44) - (10.25 \pm 0.18)\log_{10}(W)
\end{split}
\end{equation}

with associated total scatters of $\sigma_{W1} = 0.67$ and $\sigma_{W2} = 0.70$ respectively. 

These relations have two important differences from the TF relations fit using the total apparent magnitudes. First, the slopes of the TF relations in both the W1 and W2 bands are steeper when calibrated using isophotal magnitudes. This is most likely because the difference between the isophotal and total apparent magnitudes is largest for low surface brightness galaxies, which tend to be smaller galaxies with lower rotational velocities. Hence, the low rotational velocity region of the TF relation tends to be brighter when using the total apparent magnitudes, whereas the galaxies in the high rotational velocity region of the relation have more similar apparent magnitudes using both measurements. 

Second, the total scatter around the isophotal TF relation is larger by 0.01  mags in the both W1 and W2 bands. Again, this is only a marginal difference, which is most likely because the isophotal magnitudes do not include the contributions from objects located at the edges of the galaxies, and are therefore less strongly correlated with the rotational velocities, as discussed in section~\ref{Section: Data}. Hence, we conclude that calibrating the TF relation using the total apparent magnitudes is both more physically accurate and has a slightly smaller total scatter than the TF relation calibrated using isophotal magnitudes.

\section{Comparison with Near-Infrared TF Relations}\label{Section: Galaxy Evolution}
We have also compared the TF relations that we have derived in the mid-infrared with those found in the near-infrared and long-wavelength optical I \citep{K_Masters2006}, J, H and K \citep{K_Masters2008} bands;

% \begin{enumerate}
%     \item I-band: $-20.485 \pm 0.06 - (7.85 \pm 0.1)(\log W - 2.5)$
%     \item J-band: $−21.370 \pm 0.018  -(10.612 \pm 0.124)(\log(W) - 2.5)$
%     \item H-band: $−21.951 \pm 0.017 - (10.648 \pm 0.113)(\log(W) - 2.5)$
%     \item K-band: $−22.188\pm 0.015 -  (10.736 \pm 0.100)(\log(W) - 2.5).$
% \end{enumerate}

% \begin{tabular}{p{0.1cm}p{0.1cm}p{0.1cm}p{0.1cm}}
% \text{I-Band} :& M_{I} &=& (-20.485 \pm 0.06) - (7.85 \pm 0.1)(\log(W) - 2.5)\\
% \text{J-Band} :& M_{I} &=& (-21.370 \pm 0.018) - (10.612 \pm 0.124)(\log(W) - 2.5)\\
% \text{H-Band} :& M_{I} &=& (-21.951 \pm 0.017) - (10.648 \pm 0.113)(\log(W) - 2.5)\\
% \text{K-Band} :& M_{I} &=& (-22.188 \pm 0.015) - (10.736 \pm 0.100)(\log(W) - 2.5)\\
% \end{tabular}

\begin{equation}
\begin{split}
\text{I-Band} :& M_{I} = (-20.485 \pm 0.06) - (7.85 \pm 0.1)(\log(W) - 2.5)\\
\text{J-Band} :& M_{J} = (-21.370 \pm 0.018) - (10.612 \pm 0.124)(\log(W) - 2.5)\\
\text{H-Band} :& M_{H} = (-21.951 \pm 0.017) - (10.648 \pm 0.113)(\log(W) - 2.5)\\
\text{K-Band} :& M_{K} = (-22.188 \pm 0.015) - (10.736 \pm 0.100)(\log(W) - 2.5)\\
\end{split}
\end{equation}

\begin{figure}
    \centering
    \includegraphics[scale=0.45]{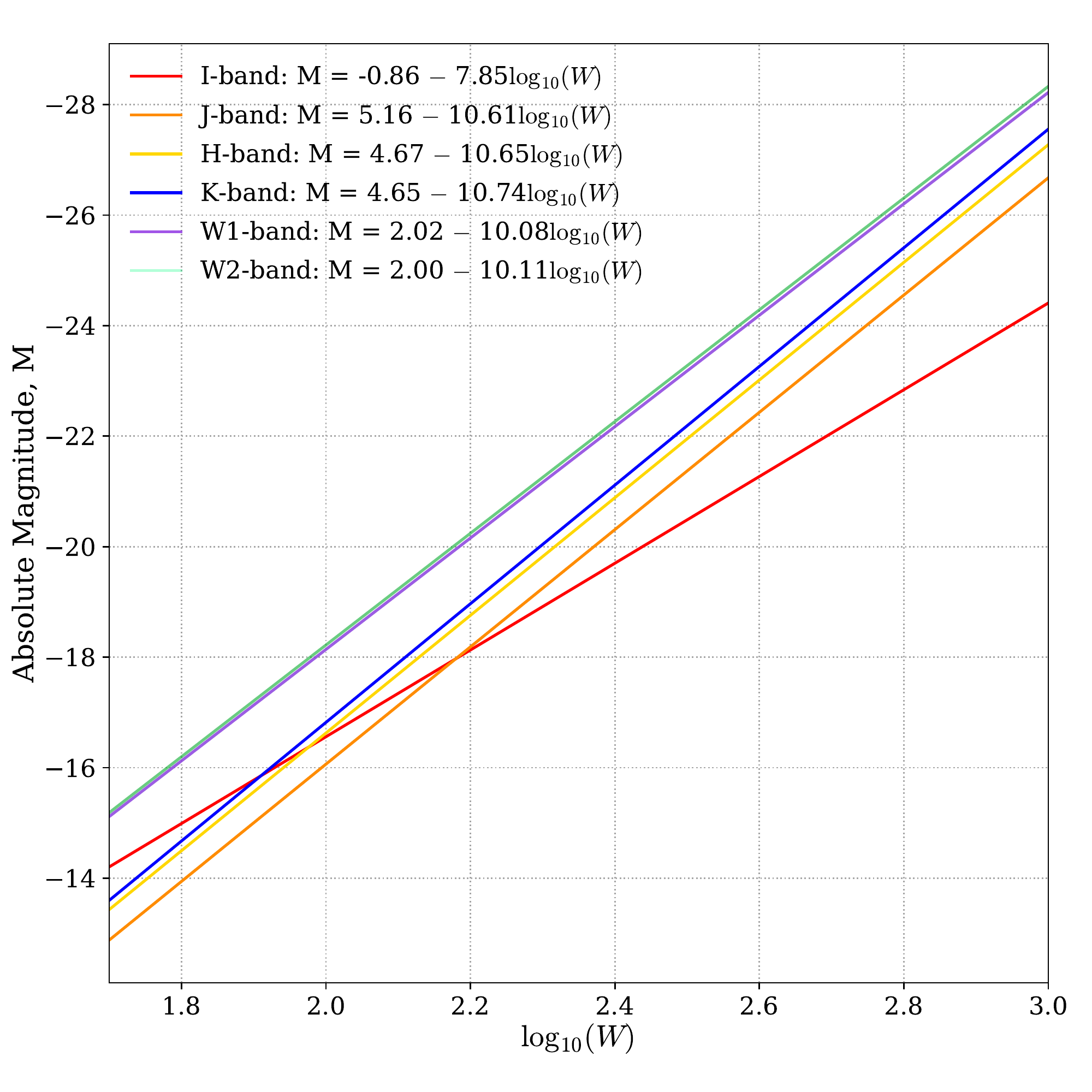}
    \caption[Comparison of the \TF relations in various infrared bands]{A comparison of the TF relations in the I, J, H, K, W1 and W2 bands, which were calibrated using the same techniques. The major difference between the J, H, K and W1 bands is an additive offset (although there are also slight variations in the slopes). The I-band is significantly less steep, and the W2 band, while being very similar to the W1 band, is slightly steeper overall and has a dimmer zero-point, comparable with the zero-points of the H and K bands.}
    \label{Fig:TF_Comparisons}
\end{figure}

Fig.~\ref{Fig:TF_Comparisons}, shows a comparison of these four relations with the relations in the W1 and W2 bands, for $W_{F50}$ measurements between 1.8 and 3. 

From this figure, it is evident that the primary difference between relations in the mid and near infrared bands is an additive offset, although the mid-infrared relations are also slightly less steep. Interestingly, the W1 and W2 band relations are almost identical within the range of rotational velocities corresponding to physical spiral galaxies.. The I-band relation, however, is significantly less steep than either the mid or near infrared relations.

There are a number of reasons why the mid-infrared relations have a large additive offset when compared with the near-infrared relations. First, the mid-infrared bands are more sensitive to low-surface brightness objects, allowing them to capture a larger amount of light from the galaxies in the sample. Moreover, while both the mid and near infrared bands trace the older stellar populations of the galaxies, the mid-infrared bands are generally able to trace the stellar emissions of both the bulge and a large fraction of the galactic disk, whereas the near-infrared predominantly capture light from the galactic bulge. Finally, the mid-infrared bands capture light from both stellar emissions and the hot gas components of the galaxies. All of these factors contribute to the mid-infrared band apparent magnitudes being systematically brighter than those in the near-infrared bands. 

However, because the W1 and W2 bands trace a larger extent of the disk than the J, H and K bands, they are also more strongly impacted by the distribution of different morphological types within the calibration sample. The low rotational velocity region of the TF relation is dominated by Sc type galaxies, which tend to have a smaller bulge component than the Sa and Sb type galaxies that dominate at high rotational velocities. Therefore, when viewed in the W1 and W2 bands, which are more sensitive to the light emitted from the galactic disk, these low-rotational velocity Sc galaxies will appear brighter  relative to the full sample, than they would when viewed in the near-infrared bands, which do not capture as much light emitted by the disk. This results in the relations in the W1 and W2 bands being slightly less steep, as can be seen in Fig.~\ref{Fig:TF_Comparisons}. 

The I-band relation is less steep than those in the infrared for two main reasons. First, the I-band is more susceptible to galactic and internal extinction than the mid-infrared bands are. Second, this band primarily captures light from the younger stellar populations in the galaxies, which tend to be more significant in disk dominated, low rotational velocity galaxies.

\section{Discussion}\label{Section: Discussion}

In this paper, we have calibrated the \TF relation in the WISE W1 ($3.4\mu$m) and W2  ($4.6\mu$m) mid-infrared bands using large samples of 848 galaxies and 857 galaxies from 31 clusters respectively. The final, fully-calibrated TF relations are 

\begin{equation}\label{FULL_TF Relations_Fin}
\begin{split}
    M_{\rm Tot, W1} &= (2.02 \pm 0.44) - (10.08 \pm 0.17)\log_{10}(W)\\
    M_{\rm Tot, W2} &= (2.00 \pm 0.44) - (10.11 \pm 0.17)\log_{10}(W),
\end{split}
\end{equation}
with the total scatter around each of these relations being $\sigma_{W1} = 0.66$ and $\sigma_{W2} = 0.69$, respectively.

\subsection{Comparison with Previous Work}
Table~\ref{Table} provides a comparison of the results derived in this paper with the parameters derived in previous calibrations of the TF relation in the mid-infrared W1 and W2 bands. 

\begin{table*}\centering
\small
\renewcommand\arraystretch{1.8}
\begin{tabular}{m{3cm}m{1.5cm}m{4cm}m{2.3cm}m{2.3cm}m{1cm}}
\Xhline{2\arrayrulewidth} 
&   \makecell[cc]{Fitting \\ Method} & \makecell[cc]{Corrections} & \makecell[cc]{Slope} & \makecell[cc]{Intercept} & \makecell[cc]{Total \\ Scatter} \\
\Xhline{2\arrayrulewidth} 
\vspace{0.1cm}
\makecell[cc]{W1 Band ($3.4\mu$m) \\ This Work} & \makecell[cc]{bivariate \\ bivariate \\ bivariate} & \makecell[cc]{- \\ Incompleteness \\ Incompleteness \& Morphology} & \makecell[cc]{$-7.94 \pm 0.17$ \\ $-9.23 \pm 0.18$ \\ $-10.08\pm 0.17$} & \makecell[cc]{$-3.38 \pm 0.42$ \\ $-0.08\pm 0.45$ \\ $2.02\pm 0.44$} & \makecell[cc]{0.77 \\ 0.68 \\ 0.67} \\
\arrayrulecolor{gray}\hline 
\makecell[cc]{W1 Band ($3.4\mu$m) \\ \citet{E_Kourkchi2020a}} & \makecell[cc]{inverse \\ inverse} & \makecell[cc]{- \\ Colour} & \makecell[cc]{$-9.47 \pm 0.14$ \\ $-9.12 \pm 0.13$} & \makecell[cc]{$3.33\pm 0.06$ \\ $2.51\pm 0.08$} & \makecell[cc]{0.58 \\ 0.47} \\
\arrayrulecolor{gray}\hline 
\makecell[cc]{W1 Band ($3.4\mu$m) \\ \citet{J_Neill2014}}& \makecell[cc]{inverse \\ inverse}  & \makecell[cc]{- \\ Colour} & \makecell[cc]{$-9.56 \pm 0.12$ \\ $-9.12 \pm 0.13$} & \makecell[cc]{$3.55\pm 0.07$ \\ $2.51\pm 0.08$} & \makecell[cc]{0.54 \\ 0.45} \\
\hline
\makecell[cc]{W1 Band ($3.4\mu$m) \\ \citet{D_Lagattuta2013}}& \makecell[cc]{forwards \\ forwards}  & \makecell[cc]{- \\ Morphology} & \makecell[cc]{$-8.13 \pm 0.6$ \\ $-10.05 \pm 0.22$} & \makecell[cc]{$2.02\pm 1.8$ \\ $2.89\pm 1.8$} & \makecell[cc]{- \\ 0.686} \\
\hline
\makecell[cc]{Spitzer $3.3\mu$m Band) \\ \citet{J_Sorce2013}}& \makecell[cc]{inverse \\ inverse}  & \makecell[cc]{- \\ Colour} & \makecell[cc]{$-9.74 \pm 0.22$ \\ $-9.13 \pm 0.22$} & \makecell[cc]{$4.01\pm 0.10$ \\ $2.49\pm 0.08$} & \makecell[cc]{0.49 \\ 0.41} \\
\Xhline{1\arrayrulewidth} 
\vspace{0.1cm}
\makecell[cc]{W2 Band ($4.6\mu$m) \\ This Work} & \makecell[cc]{bivariate \\ bivariate \\ bivariate} & \makecell[cc]{- \\ Incompleteness \\ Incompleteness \& Morphology} & \makecell[cc]{$-7.83 \pm 0.17$ \\ $-8.97 \pm 0.18$ \\ $-10.11\pm 0.17$} & \makecell[cc]{$-3.73 \pm 0.43$ \\ $-0.79\pm 0.45$ \\ $2.00\pm 0.44$} & \makecell[cc]{0.77 \\ 0.70 \\ 0.69} \\
\arrayrulecolor{gray}\hline
\makecell[cc]{W2 Band ($4.6\mu$m) \\ \citet{E_Kourkchi2020a}}& \makecell[cc]{inverse \\ inverse}  & \makecell[cc]{- \\ Colour} & \makecell[cc]{$-9.66 \pm 0.15$ \\ $-9.18 \pm 0.13$} & \makecell[cc]{$4.4\pm 0.08$ \\ $3.33\pm 0.07$} & \makecell[cc]{0.58 \\ 0.47} \\
\hline 
\makecell[cc]{W2 Band ($4.6\mu$m) \\ \citet{J_Neill2014}}& \makecell[cc]{inverse \\ inverse}  & \makecell[cc]{- \\ Colour} & \makecell[cc]{$-9.74 \pm 0.12$ \\ $-9.11 \pm 0.13$} & \makecell[cc]{$4.59\pm 0.08$ \\ $2.51\pm 0.08$} & \makecell[cc]{0.56 \\ 0.49} \\
\Xhline{2\arrayrulewidth} 
\end{tabular}
\caption[Literature values of the Tully-Fisher relation.]{Literature values for the slope and intercept of the Tully-Fisher relation in the mid-infrared bands for comparison with our calibration. The intercepts of the TF relations in these papers have been adjusted in this table to account for the use of $\log_{10}(W) - 2.5$ as the 'x' variable in these papers.} 
\label{Table}
\end{table*}

The TF parameters derived in this paper exhibit differences from those derived in previous calibrations in the W1 and W2 bands. Of the results presented in table~\ref{Table}, the calibrations performed by \citet{J_Sorce2013}, \cite{J_Neill2014} and \citet{E_Kourkchi2020a} [1 - 3] employed the \textit{inverse} TF relation to derive the TF slope and intercept. It is well established that statistical biases in TF calibration samples have a larger impact on the forwards and bivariate fitting techniques for the TF relation than they do on the inverse fitting method \citep{J_Willick1994}. Hence, the inverse TF relation is often employed to avoid the use of bias-correcting techniques. The slopes without bias corrections from these three papers closely align with the slope in the present work after the incompleteness bias correction ($b_{\rm debiased} = -9.23 \pm 0.18$).

However these three calibrations then performed a colour correction rather than a morphological type correction. Colour corrections provide an approximate continuous parameterisation of morphological type corrections as Sa and Sb type galaxies generally appear more red than Sc type galaxies due to their older stellar populations, which are stronger emitters in the long-wavelength optical and near-infrared band \citep{J_Sorce2013}. In the colour corrections applied in [1-3], the `bluer' galaxies in the sample, which generally correspond to Sc morphological types, were shifted downwards to match the distribution of the `redder' Sa and Sb galaxies. This is the opposite to the morphological type correction applied in this paper, where the Sa and Sb type galaxies were shifted upwards to more closely match the distribution of Sc galaxies. Both corrections changed the slopes of the calibrated TF relations by approximately 0.5 mags but in opposite directions. 

The fourth calibration in the W1 band, performed by \citet{D_Lagattuta2013}, used a forwards fitting technique, and performed a direct morphological type correction, which is more similar to the technique used in this paper. Their initial slope of $b = -8.13 \pm 0.6$ is close to the initial slope of $b = -7.94 \pm 0.14$ derived in this paper using a bivariate fit to the raw data in the W1 band. The morphological correction applied by \citet{D_Lagattuta2013} increased the slope of the TF relation to a value similar to the fully-corrected slope in this paper. Their morphological correction was significantly larger than the morphological correction in this paper. This is likely because Sa and Sb type galaxies tend to suffer from higher levels of incompleteness, and their morphological correction would have partially incorporated a bias correction. 

\subsection{Intrinsic Scatter around the TF Relation}
One of the primary aims of this paper was characterise the intrinsic scatter around the TF relation in the mid-infrared bands. The primary advantage of using mid-infrared photometry in the calibration of the TF relation is the increased sensitivity to low surface brightness galaxies, which allows for WISE photometry to capture the emissions from the disks of spiral galaxies out to much larger radii than the photometry performed at the near-infrared or optical wavelengths. This is ideal for calibrating the TF relation, as the mid-infrared apparent magnitude measurements should be more reflective of the \textit{total} mass of the old stellar populations in the galaxies, rather than just the stars in the inner regions of the disk and the bulge. However, while this could in principle, lead to a tighter TF correlation, it may also introduce additional scatter into the absolute magnitude measurements due to differences in the populations.

The disks of spiral galaxies are inherently more variable than their bulges. Active star formation is mostly confined to the disk, so the properties of the disk reflect the evolutionary history of the galaxy. This means additional factors such as age, external environment, metallicity and merger history of the galaxy will likely have a larger impact on the total luminosity of the disk than on the luminosity of the bulge. 

In addition, light in the mid-infrared W1 and W2 bands includes emissions from hot inter-stellar dust in the disk, polycyclic aromatic hydrocarbons (PAHs), and active galactic nuclei \citep{S_Meidt2012}, whereas light in the near-infrared K-band is completely dominated by the light from old stars mostly found in the bulge and inner regions of the disk. Studies have found that as much as 12\% of a spiral galaxy's emissions in the mid-infrared are produced by hot inter-stellar gas, with this value increasing further in the W2 band \citep{S_Meidt2012, J_Sorce2013}. The spatial distribution and density of the dust in the disk of a spiral galaxy is much more variable than the distribution and density of the old stellar population \citep{J_Neill2014}. Hence, it is likely that these hot dust emissions in the mid-infrared are one of the main causes of the increased scatter around the TF relations in these bands. Moreover, these emissions are primarily produced by dust that has been thermally heated by very young, hot stars, meaning that the hot dust emissions effectively trace the regions of star formation in spiral galaxies, further increasing the variability between different galaxies. 

It is also important to note that by capturing the light emitted by the hot dust in the galaxy, the W1 and W2 bands are tracing a component of the galaxy with a significantly higher mass-to-light ratio than the stellar population. This is the most likely cause for the slightly flatter TF slopes derived in these bands compared to the near-infrared, as well as the flattening of the slope in the W2 band compared to the W1 band.

Finally, up to 10\% of the spiral galaxies in the local universe have been shown to have active galactic nuclei \citep{Brinchmann2004}. The mid-infrared radiation produced by these regions of the galaxies can potentially dwarf their stellar emissions, further increasing the scatter around the TF relations. 

\subsection{Implications of using Cluster Samples}
The advantages of cluster samples are that they allow for the use of a larger number of galaxies and the bias correction technique is more straightforward than for field galaxies. However, a disadvantage is that galaxies from clusters are a sub-set of spiral galaxies and might not accurately reflect the intrinsic TF relation for spiral galaxies.  

The external environment of a cluster galaxy is very different from that of a field galaxy. The space between galaxies within clusters is generally filled with a hot medium consisting of gas and dust. This material exerts a pressure on the galaxies (called ram pressure) which can apply strong enough forces to strip the cool, dense gas required for star formation out of the galaxies moving through the intracluster medium. However, the ram pressure can also compress the gas within galaxies, leading to accelerated star formation. These effects may produce differences between cluster populations of spiral galaxies and field galaxies which result in different TF relation for the two populations. 

Any differences could also affect the calculation of the incompleteness bias, as it uses a luminosity functions for spiral galaxies derived from combined cluster and field samples. The shape of the luminosity function is one of the most important factors in the bias correction procedure, and may significantly alter the results of the calibration. In addition, our calculations of completeness functions for some clusters in this paper show that while the absolute magnitude at which the cluster is 50\% complete appears to have a logarithmic dependence on the distance to the cluster, the steepness of the completeness function does not. This suggests that the properties of different galaxy clusters are highly variable, and may differ significantly depending on factors such as their location, size and density, and evolutionary history. Hence, the use of cluster galaxies in the calibration of the TF relation may introduce an additional level of variability, thereby increasing scatter in the relation. However, this effect is partially mitigated within this calibration by the fact that our sample contains both \textit{bona fide} cluster members as well as galaxies in the vicinity of the clusters (within 1 and 2 Abell radii).

\section*{Acknowledgements}

% The Acknowledgements section is not numbered. Here you can thank helpful
% colleagues, acknowledge funding agencies, telescopes and facilities used etc.
% Try to keep it short.

This research was funded in part by the Australian Government through the Australian Research Council project number FL180100168. 

We would like to thank Dr Cullan Howlett for providing useful comments during the course of this research.  

%%%%%%%%%%%%%%%%%%%%%%%%%%%%%%%%%%%%%%%%%%%%%%%%%%
\section*{Data Availability}
All codes are available via GitHub, and the data will be made available on reasonable request to the corresponding author.

%%%%%%%%%%%%%%%%%%%% REFERENCES %%%%%%%%%%%%%%%%%%

% The best way to enter references is to use BibTeX:

\bibliographystyle{mnras}
\bibliography{bibliography} % if your bibtex file is called example.bib

% Alternatively you could enter them by hand, like this:
% This method is tedious and prone to error if you have lots of references
%\begin{thebibliography}{99}
%\bibitem[\protect\citeauthoryear{Author}{2012}]{Author2012}
%Author A.~N., 2013, Journal of Improbable Astronomy, 1, 1
%\bibitem[\protect\citeauthoryear{Others}{2013}]{Others2013}
%Others S., 2012, Journal of Interesting Stuff, 17, 198
%\end{thebibliography}

%%%%%%%%%%%%%%%%%%%%%%%%%%%%%%%%%%%%%%%%%%%%%%%%%%

%%%%%%%%%%%%%%%%% APPENDICES %%%%%%%%%%%%%%%%%%%%%

\appendix

% Don't change these lines
\bsp	% typesetting comment
\label{lastpage}
\end{document}